\documentclass[review]{elsarticle}

\usepackage{lineno,hyperref}
\usepackage[english]{babel}
\usepackage{amsmath}
\usepackage{amsfonts}
\usepackage{textgreek}
\usepackage{natbib}
\setcitestyle{numbers,square,sort&compress}
\usepackage{graphicx}
\usepackage{breqn}
\usepackage{undertilde}
\usepackage{appendix}
\usepackage{accents}
\usepackage{enumerate}
\usepackage{caption}
\usepackage{algorithm}
\usepackage{stmaryrd}
\usepackage{color}
\usepackage{subcaption}

\usepackage[hang,flushmargin]{footmisc} 

\modulolinenumbers[5]

\journal{IUTAM Lyngby}

\biboptions{sort&compress}

\begin{document}

\begin{frontmatter}

\title{On the competition between dislocation transmission and crack nucleation at phase boundaries {\let\thefootnote\relax\footnote{{This article was presented at the IUTAM Symposium on Size-Effects in Microstructure and Damage Evolution at Technical University of Denmark, 2018}}}}

\author[TUe]{F.~Bormann}
\author[TUe]{R.H.J.~Peerlings}
\author[TUe]{M.G.D.~Geers}

\address[TUe]{Department of Mechanical Engineering, Eindhoven University of Technology, PO Box 513, 5600 MB Eindhoven, The Netherlands}
%
%
%
%
\begin{abstract}
	The interaction of dislocations with phase boundaries is a complex phenomenon, that is far from being fully understood. A 2D Peierls-Nabarro finite element (PN-FE) model for studying edge dislocation transmission across fully coherent and non-damaging phase boundaries was recently proposed. This paper brings a new dimension to the complexity by extending the PN-FE model with a dedicated cohesive zone model for the phase boundary. With the proposed model, a natural interplay between dislocations, external boundaries and the phase boundary, including decohesion of that boundary, is provided. It allows one to study the competition between dislocation transmission and phase boundary decohesion.
	Commonly, the interface potentials required for glide plane behaviour and phase boundary decohesion are established through atomistic simulations. They are corresponding to the misfit energy intrinsic to a system of two bulks of atoms that are translated rigidly with respect to each other.
	It is shown that the blind utilisation of these potentials in zero-thickness interfaces (as used in the proposed model) may lead to a large quantitative error. Accordingly, for physical consistency, the potentials need to be reduced towards zero-thickness potentials. In this paper a linear elastic reduction is adopted.
	With the reduced potentials for the glide plane and the phase boundary, the competition between dislocation transmission and phase boundary decohesion is studied for an 8-dislocation pile-up system. Results reveal a strong influence of the phase contrast in material properties as well as the phase boundary toughness on the outcome of this competition. In the case of crack nucleation, the crack length shows an equally strong dependency on these properties.
\end{abstract}


\end{frontmatter}

%
%
%
%
\section{Introduction}
Dislocation interactions with grain and phase boundaries are known to be complex phenomena. Depending on the geometrical properties (e.g. grain misorientation) and the material properties (intra- and interphase), a variety of events may occur. To gain a more profound insight in the interplay between dislocations and internal boundaries, atomistic studies on various grain and phase boundaries have been performed \cite{VanSwygenhoven2002,DeKoning2003, Lasko2005,Wang2011a,Wang2012,Wang2015,Shimokawa2014,Sobie2014,Pan2014, Cui2014,Elzas2016}. Reported events are dislocation obstruction, dislocation reflection, dislocation nucleation, dislocation transmission across the boundary, dislocation absorption into the boundary and dislocation induced decohesion. However, the underlying mechanisms controlling these phenomena are not properly understood -- let alone their interplay and/or competition. To acquire a better understanding of the mechanics of these events, each isolated event needs to be scrutinised. Atomistic models generally are not suitable for this because they do not allow one to "switch off" certain mechanisms. Several alternative modelling approaches have been proposed to capture the local dislocation behaviour. The most common approaches are the Peierls--Nabarro (PN) model \cite{Peierls1940,Nabarro1947,Hirth1982}, phase-field based models \cite{Shen2004,Hunter2013a,Mianroodi2015} and Field Dislocation Mechanics \cite{Acharya2010,Zhang2015,Zhang2017}. Using these models, dislocation transmission across simple grain and phase boundary structures was recently studied \cite{Anderson2001,Shehadeh2007,Zeng2016,Bormann2018}.\par
Here we add a novel dimension to the problem beyond transmission, by extending the recently proposed 2D Peierls--Nabarro finite element (PN-FE) model \cite{Bormann2018} to incorporate decohesion. This extension enables us to study how the local stresses due to a dislocation or a pile-up of dislocations may result in an interface crack. In some other cases, a dislocation (of the pile-up) may be transmitted without any cracking. Our goal is to study this competition of mechanisms and the dependence of its outcome on the physical properties, e.g. phase contrast, interface properties, etc. \par
In this paper, we consider the idealised problem of a two-phase microstructure in two dimensions. It consists of a soft phase which is flanked by a harder phase. Embedded in both phases lies a single glide plane perpendicular to and continuous across the fully coherent phase boundary. Centred in the soft phase, a dislocation source is assumed that emits edge dislocation dipoles under the influence of an externally applied shear load. The glide plane is modelled in accordance with the PN model as a zero-thickness interface, splitting the microstructure into two regions of linear elasticity. Along the glide plane, an energy based interface model is employed to capture the structure and motion of dislocations. It entails a periodic, and thus non-convex, potential in terms of the relative tangential displacement, or disregistry, between the two elastic regions. Dislocation arise naturally as localised transitions from one well of this potential to the next. The phase boundary is fitted with a dedicated cohesive zone model which allows for a relative normal displacement, or opening, at the cost of an energy -- which, for large openings, approaches the fracture toughness. The total free energy, which comprises the elastic strain energy, the misfit energy of the glide plane and the cohesive energy of the phase boundary, is highly non-convex. To minimise it, the model is discretised by finite elements and solved numerically by the Truncated Newton method \cite{Bormann2018b}. \par 
While it seems intuitive to employ atomistics based potentials for the glide plane and phase boundary, such potentials correspond to a misfit energy that is intrinsic to the finite distance between two layers of atoms. When employed to a zero-thickness interface, as done in the present model, erroneous results may be obtained due to the incorporation of the (linear) elastic response between the two layers of atoms, which is in contradiction with the zero thickness of the interface models. Hence, to restore physical consistency, Rice \cite{Rice1992} and later Sun et. al \cite{Sun1993} proposed the exclusion of this linear elastic response from the atomistically calculated potentials and its reduction towards a non-linear potentials that correspond to zero-thickness interfaces. In later studies, Xu et al. \cite{Xu2000a,Xu2003} showed that the linear elastic potential reduction has a significant influence on the Peierls stress and on the activation energy for dislocation nucleation from a crack tip, and it hence may not be neglected -- as is commonly done in the literature -- including our earlier work in Reference \cite{Bormann2018}. \par
In the first part of this paper we study the influence of the linear elastic reduction on the obtained results for the interplay of dislocations with a perfectly bonded, as well as a decohering phase boundary. In the second part, the reduced potentials are employed for a parameter study on the competition between dislocation transmission and crack nucleation as well as on the resulting crack length. 
The paper is organised as follows. In Section \ref{sect:Problem_Statement} the Peierls--Nabarro cohesive zone (PN-CZ) model for dislocations interacting with a decohering phase boundary is formulated. Its capability of modelling dislocation transmission and dislocation induced interface decohesion is illustrated in Section \ref{sect:Unreduced_Results} to familiarise the reader with the mechanics of the problem at hand. Section \ref{sect:Reduced_Potential} introduces the linear elastic reduction of the corresponding potentials and demonstrates its influence on the dislocation behaviour. A parameter study on the competition between dislocation transmission and phase boundary decohesion follows in Section \ref{sect:trans_vs_dec}. Finally, conclusions are presented in Section \ref{sect:discussion}.
%
%
%
%
\section{The Peierls--Nabarro cohesive zone (PN-CZ) model}
\label{sect:Problem_Statement}
%
%
\subsection{Model formulation}
\begin{figure}[htbp]
	\centering
	\includegraphics[width=0.75\linewidth]{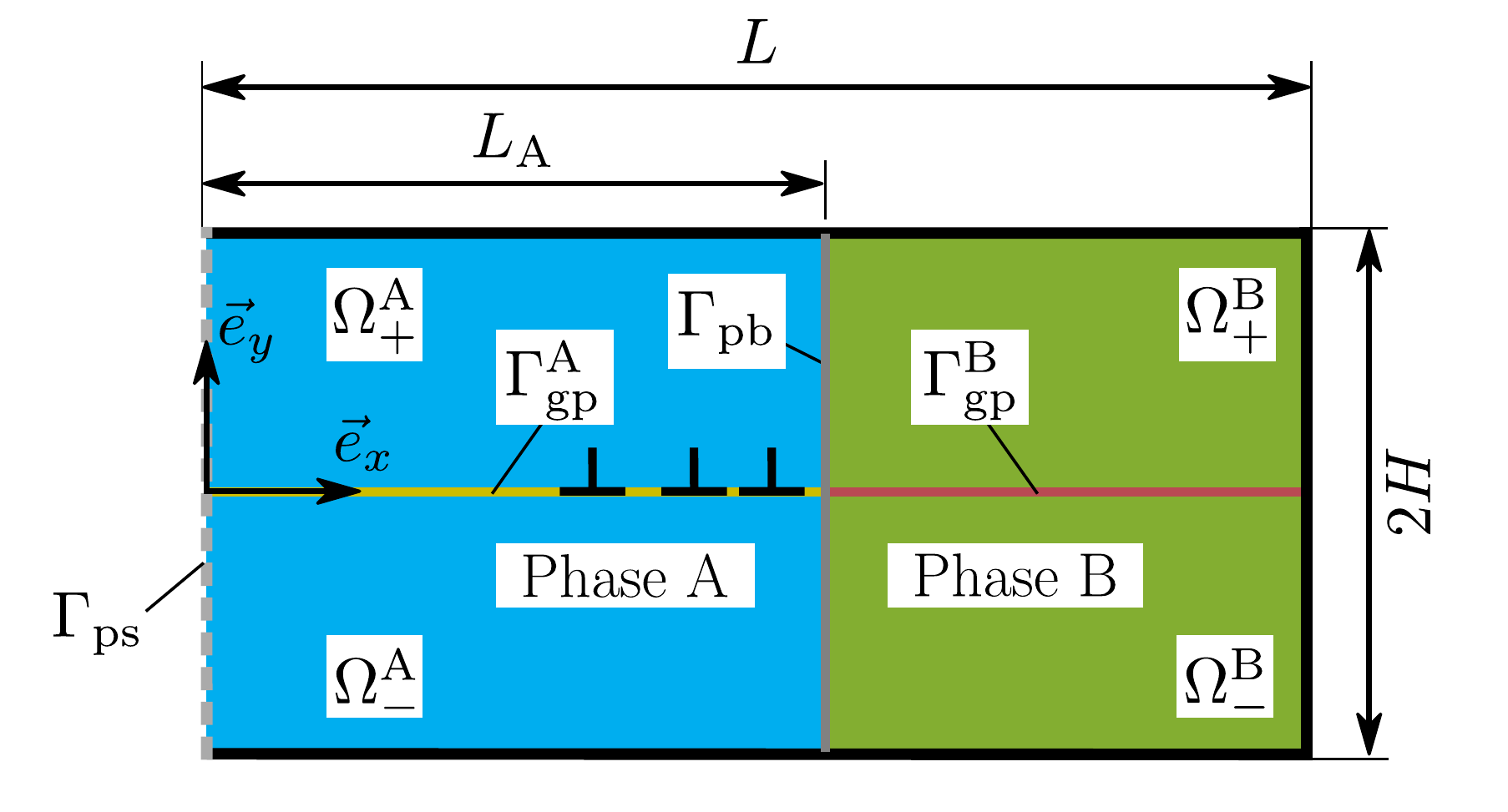}
	\caption{Continuum PN-CZ model for edge dislocation dipoles interacting with a phase boundary in a two-phase microstructure. $\Gamma_{\mathrm{ps}}$ denotes the symmetry plane of the dipole problem.}
	\label{fig:FEM-Model}
\end{figure}

Let $\Omega$ be the two-phase microstructure illustrated in Figure \ref{fig:FEM-Model}. Any material point in $\Omega$ is mapped by the position vector $\vec{x}$ in the Eucledian point space $R^2$ with basis vectors $\vec{e}_x$ and $\vec{e}_y$. The glide plane $\Gamma_{\mathrm{gp}}$ and the phase boundary $\Gamma_{\mathrm{pb}}$ are zero thickness interfaces, splitting $\Omega$ into the subdomains $\Omega_\pm^i$ with $i\in \left\{\mathrm{A},\mathrm{B}\right\}$ (see Figure \ref{fig:FEM-Model}):
\begin{align}
	\Omega =& \Omega^{\mathrm{A}} \cup \Omega^{\mathrm{B}} \\
	\Gamma_{\mathrm{gp}}=&\Gamma_{\mathrm{gp}}^{\mathrm{A}} \cup \Gamma_{\mathrm{gp}}^{\mathrm{B}}\\
	\Omega^i =& \Omega^i_+ \cup \Omega^i_-\\
	\partial\Omega =& (\partial\Omega^{\mathrm{A}}\setminus\Gamma_{\mathrm{pb}})\cup(\partial\Omega^{\mathrm{B}}\setminus\Gamma_{\mathrm{pb}})\\
	\partial\Omega^i =&
	(\partial\Omega^i_+\setminus\Gamma_{\mathrm{gp}}^i)\cup(\partial\Omega^i_-\setminus\Gamma_{\mathrm{gp}}^i)
\end{align}
For simplicity, $\Gamma_{\mathrm{gp}}$ is oriented here with its normal $\vec{e}_{n,\mathrm{gp}}=\vec{e}_y$ and its slip direction $\vec{e}_{t,\mathrm{gp}}=\vec{e}_x$; the normal of $\Gamma_{\mathrm{pb}}$ is $\vec{e}_{n,\mathrm{pb}}=\vec{e}_x$. Assuming all non-linear deformation of $\Omega$ to be confined to $\Gamma_{\mathrm{gp}}$ and $\Gamma_{\mathrm{pb}}$, the total free energy (per unit thickness out of the plane of the sketch of Figure \ref{fig:FEM-Model}) of $\Omega$ is defined as
\begin{equation}
	\label{eq:internal-energy}
	\Psi=\int_{\bar{\Omega}}\psi_{\mathrm{e}}\,\mathrm{d}\Omega+\int_{\Gamma_{\mathrm{gp}}}\psi_{\mathrm{gp}}\,\mathrm{d}\Gamma+\int_{\Gamma_{\mathrm{pb}}}\psi_{\mathrm{pb}}\,\mathrm{d}\Gamma
\end{equation}
with $\bar{\Omega} = \Omega\setminus(\Gamma_{\mathrm{gp}}\cup\Gamma_{\mathrm{pb}})$. Here, $\psi_{\mathrm{e}}$ is the elastic strain energy density in $\Omega^{\mathrm{A}}_\pm$ and $\Omega^{\mathrm{B}}_\pm$, calculated by standard linear elasticity under a plane strain condition; $\psi_{\mathrm{gp}}$ is the glide plane potential describing the misfit energy density along $\Gamma_{\mathrm{gp}}^{\mathrm{A}}$ and $\Gamma_{\mathrm{gp}}^{\mathrm{B}}$; $\psi_{\mathrm{pb}}$ is the phase boundary potential defining the reversible cohesive energy density along $\Gamma_{\mathrm{pb}}$. Phase specific material properties apply for $\psi_{\mathrm{e}}$ and $\psi_{\mathrm{gp}}$.\par
Both interface potentials, $\psi_{\mathrm{gp}}$ and $\psi_{\mathrm{pb}}$, are functions of the relative displacement between initially coinciding points on $\Gamma_{\mathrm{gp}}$ and $\Gamma_{\mathrm{pb}}$, respectively:
\begin{align}
	\label{eq:gp-disregistry}
	\vec{\Delta}_{\mathrm{gp}} = [[\vec{u}]] = &\vec{u}_+-\vec{u}_-,\qquad \vec{x}\in\Gamma_{\mathrm{gp}}\\
	\label{eq:pb-disregistry}
	\vec{\Delta}_{\mathrm{pb}} = [[\vec{u}]] = &
	\vec{u}^{\,\mathrm{B}}-\vec{u}^{\,\mathrm{A}},\qquad \vec{x}\in\Gamma_{\mathrm{pb}}
\end{align}
Due to the alignment of $\Gamma_{\mathrm{gp}}$ with the global basis vectors, the tangential relative displacement, or disregistry, of the glide plane is defined as $\Delta_{\mathrm{gp}}=\vec{\Delta}_{\mathrm{gp}}\cdot\vec{e}_x$; the normal relative displacement, or opening, of the phase boundary is $\Delta_{\mathrm{pb}}=\vec{\Delta}_{\mathrm{pb}}\cdot\vec{e}_x$. \par
\begin{figure}[htbp]
	\centering
	\includegraphics[width=1.\linewidth]{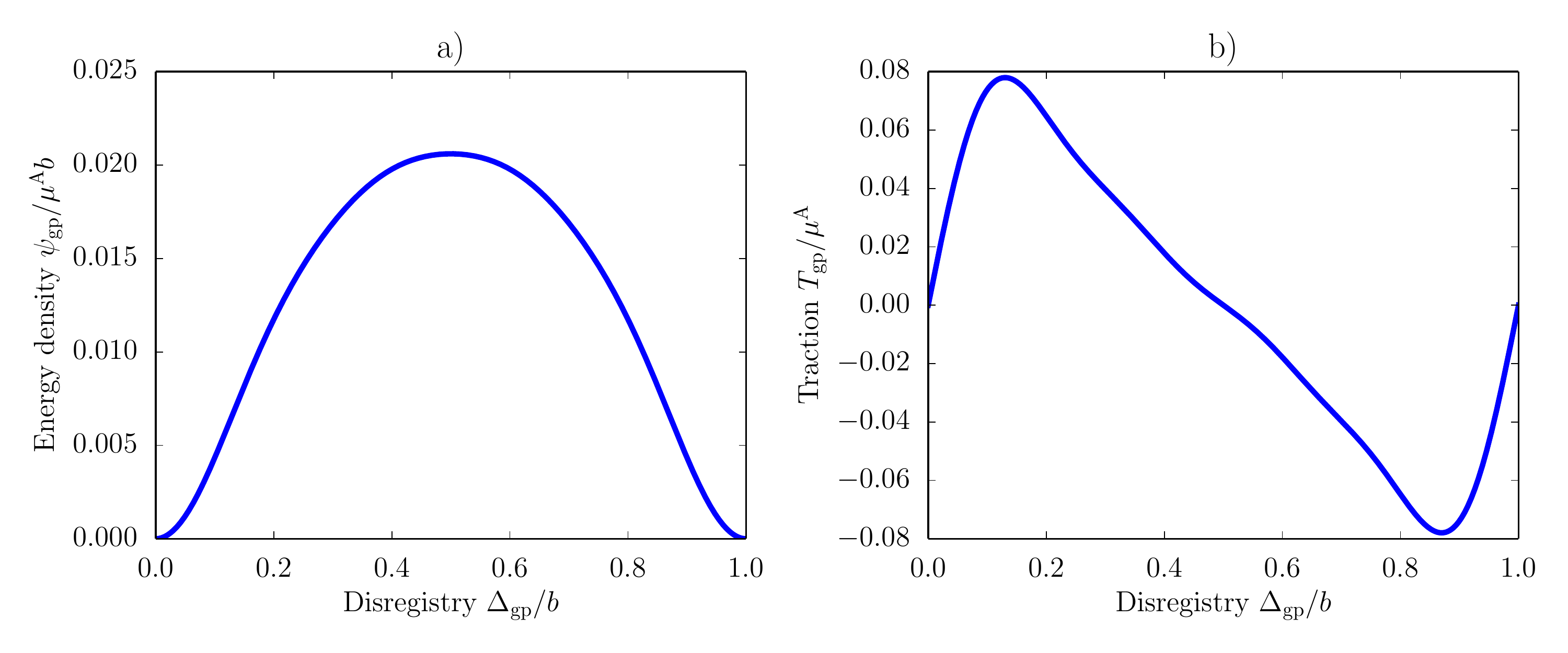}
	\caption{(a) Glide plane energy density $\psi_{\mathrm{gp}}$ and (b) glide plane traction $T_{\mathrm{gp}}$ as a function of the disregistry $\Delta_{\mathrm{gp}}$.}
	\label{fig:unreduced_GP_energy_traction}
\end{figure}

In this paper, a Fourier based glide plane potential is employed \cite{Bormann2018c}:
\begin{equation}
	\label{eq:original_gp-law}
	\psi_{\mathrm{gp}}(\Delta_{\mathrm{gp}}) = \sum_{k}\frac{1}{k}\gamma_{\mathrm{us},k}^i\sin^2\left(\frac{k\pi\Delta_{\mathrm{gp}}}{b^i}\right)
\end{equation}
where $\gamma_{\mathrm{us},k}^i$ are the Fourier parameters and $b^i$ the magnitude of the Burgers vector associated with Phase $i$. Any normal relative displacement $\vec{\Delta}_{\mathrm{gp}}\cdot\vec{e}_y$ along the glide plane is constrained to zero. The glide plane tractions are given by
\begin{equation}
	\label{eq:gp-tractions}
	T_{\mathrm{gp}} =\frac{\mathrm{d}\psi_{\mathrm{gp}}}{\mathrm{d}\Delta_{\mathrm{gp}}}
\end{equation}
The glide plane energy density $\psi_{\mathrm{gp}}$ and the glide plane traction $T_{\mathrm{gp}}$ are plotted in Figure \ref{fig:unreduced_GP_energy_traction} as a function of $\Delta_{\mathrm{gp}}$, for the parameters specified in Section \ref{sect:Parameters_set}.\par
\begin{figure}[htbp]
	\centering
	\includegraphics[width=1.\linewidth]{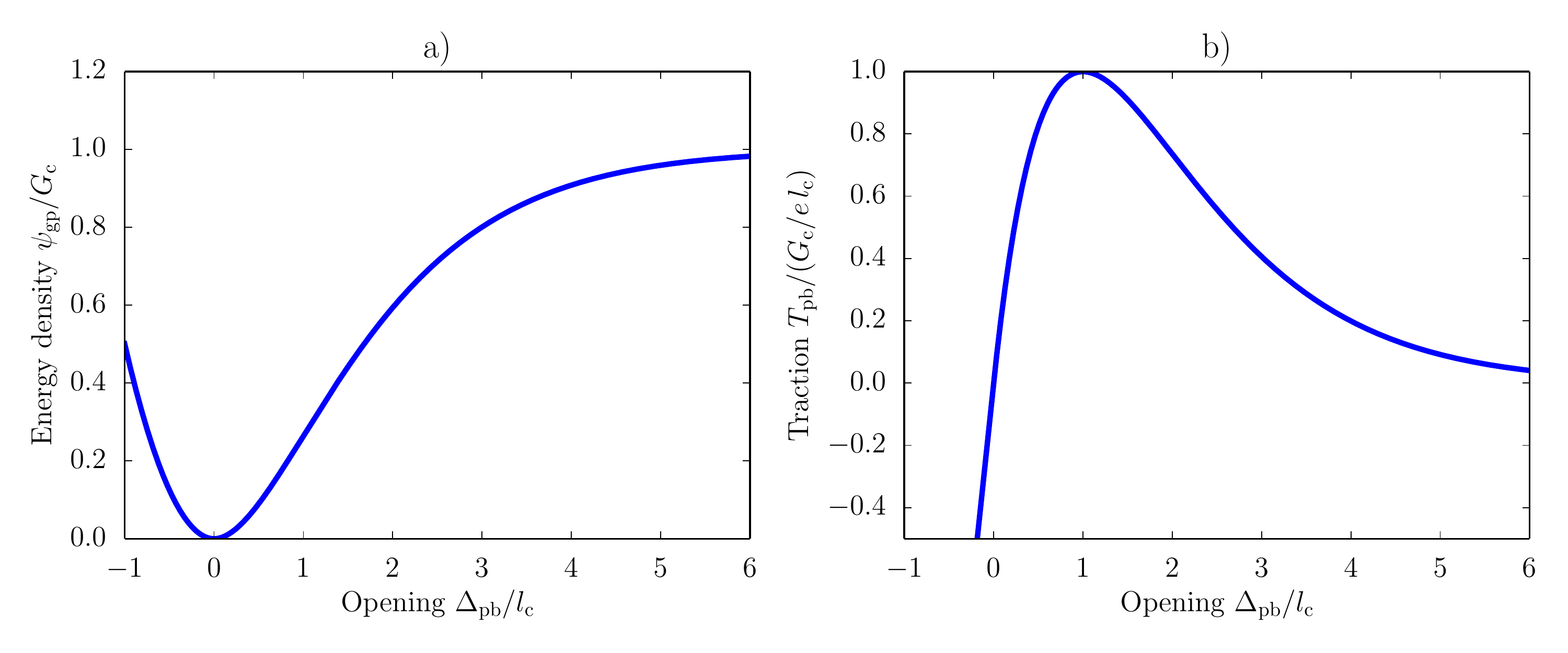}
	\caption{(a) Phase boundary energy density $\psi_{\mathrm{pb}}$ and (b) phase boundary traction $T_{\mathrm{pb}}$ as a function of the opening $\Delta_{\mathrm{pb}}$.}
	\label{fig:unreduced_PB_energy_traction}
\end{figure}

The phase boundary potential adopted here is a modified version of the Rose--Ferrante--Smith universal binding relation \cite{Rose1983}. In it, the exponential behaviour is replaced by a quadratic expression in the compressive regime to facilitate the linear elastic reduction introduced in Section \ref{sect:Reduced_Potential}. The phase boundary potential reads
\begin{equation}
	\label{eq:original_pb-law}
	\psi_{\mathrm{pb}}(\Delta_{\mathrm{pb}}) =
	\begin{cases}
		G_{\mathrm{c}}\left[1-\left[1+\frac{\Delta_{\mathrm{pb}}}{l_{\mathrm{c}}}\right]\exp\left(-\frac{\Delta_{\mathrm{pb}}}{l_{\mathrm{c}}}\right)\right],& \Delta_{\mathrm{pb}}\ge 0\\
		\frac{1}{2}G_{\mathrm{c}}\left(\frac{\Delta_{\mathrm{pb}}}{l_{\mathrm{c}}}\right)^2,& \Delta_{\mathrm{pb}}<0
	\end{cases}
\end{equation}
with the work of separation $G_{\mathrm{c}}$ and the characteristic length $l_{\mathrm{c}}$, defined as the opening where $\partial^2\psi_{\mathrm{pb}}/\partial\Delta_{\mathrm{pb}}^2=0$. The tangential sliding $\Delta_{\mathrm{pb}}=\vec{\Delta}_{\mathrm{pb}}\cdot\vec{e}_y$ of the phase boundary is constrained to zero. The phase boundary tractions read
\begin{equation}
	\label{eq:pb-tractions}
	T_{\mathrm{pb}}=\frac{\mathrm{d}\psi_{\mathrm{pb}}}{\mathrm{d}\Delta_{\mathrm{pb}}}
\end{equation}
The phase boundary energy density $\psi_{\mathrm{pb}}$ and the phase boundary traction $T_{\mathrm{pb}}$ are illustrated in Figure \ref{fig:unreduced_PB_energy_traction} as a function of $\Delta_{\mathrm{pb}}$.
%
%
\subsection{Boundary conditions}
Edge dislocations present in the domain $\Omega$ as sketched in Figure \ref{fig:FEM-Model} are thought of as a part of an edge dislocation dipole centred at $\vec{x}=\vec{0}$. They are subjected to an externally applied shear deformation. Together, these assumptions give rise to the symmetry boundary condition $\vec{u}(y) = -\vec{u}(-y)$ on the vertical symmetry plane $\Gamma_{\mathrm{ps}}$ with
\begin{equation}
	\Gamma_{\mathrm{ps}}= \{(0,y)| -H<y<H\}
\end{equation} 
For conciseness, the term dislocation dipole will be replaced in the following by dislocation whenever this does not lead to confusion.\par
On the outer boundary $\partial\Omega\setminus\Gamma_{\mathrm{ps}}$ a shear deformation is imposed, which, for a linear elastic model response (no glide plane), induces a constant shear stress $\tau=t\bar{\tau}$ in $\Omega$. Here, $\bar{\tau}$ is the target shear load and $t\in\left[0,1\right]$ a pseudo-time to capture the model's evolution under an increasing shear load. The corresponding Dirichlet boundary conditions read
\begin{align}
	\label{eq:boundary_value1}
	\vec{u}&=\frac{\bar{\tau}}{\mu^{\mathrm{A}}} x\,t\,\vec{e}_y && \quad\mathrm{on}\:\partial\Omega^{\mathrm{A}}\setminus\left(\Gamma_{\mathrm{pb}}\cup\Gamma_{\mathrm{ps}}\right)\\
	\label{eq:boundary_value2}
	\vec{u}&=\frac{\bar{\tau}}{\mu^{\mathrm{B}}}\left[x-\left(1-\frac{\mu^{\mathrm{B}}}{\mu^{\mathrm{A}}}\right)L_{\mathrm{A}}\right]\,t\,\vec{e}_y && \quad\mathrm{on}\:\partial\Omega^{\mathrm{B}}\setminus\Gamma_{\mathrm{pb}}
\end{align}
with $\mu^i$ as the shear modulus of Phase $i$.
%
%
\subsection{Solution method}
For the evaluation of the PN-CZ model under the applied boundary conditions \eqref{eq:boundary_value1},\eqref{eq:boundary_value2} at time $t_n$, the non-convex total free energy of Eq. \eqref{eq:internal-energy} needs to be minimised. To this end, the full problem is discretised by finite elements and solved with the adapted truncated Newton method, as outlined in \cite{Bormann2018b}. To nucleate dislocations, i.e. no annihilation of the dipole occurs, the methodology outlined in \cite{Bormann2018} is followed.
\subsection{Parameter set used}
\label{sect:Parameters_set}
In the analyses presented in this paper, the material properties of Phase A, i.e. elasticity parameters and glide plane properties, are chosen consistently with molecular statics results for a 2D hexagonal lattice \cite{Bormann2018c}. All parameters are parametrised with respect to the shear modulus $\mu^{\mathrm{A}}$ and the Burgers vector $b^{\mathrm{A}}=b$. Poisson's ratio is defined as $\nu^{\mathrm{A}}=0.25$ and the Fourier parameters for the glide plane are taken as listed in Table \ref{tab:Phase_A}. The material parameters of Phase B are defined through the phase contrast $k_{\mathrm{m}}$ as $\mu^{\mathrm{B}}=k_{\mathrm{m}}\mu^{\mathrm{A}}$ and $\gamma_{\mathrm{us},k}^{\mathrm{B}}=k_{\mathrm{m}}\gamma_{\mathrm{us},k}^{\mathrm{A}}$; a homogeneous Poisson's ratio applies, i.e. $\nu^{\mathrm{B}}=\nu^{\mathrm{A}}=\nu$. The coherent phase boundary implies $b^{\mathrm{B}}=b^{\mathrm{A}}=b$. The phase boundary properties have also been calibrated on molecular statics results and are defined as $l_{\mathrm{c}}=0.14\,b$ and $G_{\mathrm{c}}=k_{\mathrm{pb}}\left(1+k_{\mathrm{m}}\right)G_{\mathrm{c},0}$ with $G_{\mathrm{c},0}=7.24 \cdot 10^{-2}\mu^{\mathrm{A}}b$ and the toughness factor $k_{\mathrm{pb}}$, which allows one to vary the phase boundary toughness and strength simultaneously. The model dimensions are chosen as $L_{\mathrm{A}}=2000\,b$, $L=3000\,b$ and $H=2250\,b$. The full model is discretised by linear triangular elements with one central Gauss point, for $\Omega^i_\pm$, and by linear interface elements with two Gauss points, for $\Gamma_{\mathrm{gp}}$ and $\Gamma_{\mathrm{pb}}$. A minimum element size of $b/8$ is adopted to adequately capture the dislocation behaviour and phase boundary decohesion. Outside of the region of interest, the mesh coarsens rapidly.\par
For the load application, a target shear load of $\bar{\tau}=0.07\,\mu^{\mathrm{A}}$ is considered, which refers to $90\% $ of the glide plane traction amplitude $\max\left\{T_{\mathrm{gp}}\right\}$ of Phase~$\mathrm{A}$. Note that this rather large target stress is solely chosen for the purpose of a qualitative study. Results are to be interpreted carefully in the context of the adopted small strain framework. 
\begin{table}[htbp]
	\caption{Fourier parameters for the glide plane potential of Phase A.}
	\centering
	\begin{tabular}{c|cccc}
		Parameter	&	$\gamma_{\mathrm{us},1}/\mu^{\mathrm{A}}b$	&	$\gamma_{\mathrm{us},2}/\mu^{\mathrm{A}}b$	&	$\gamma_{\mathrm{us},3}/\mu^{\mathrm{A}}b$	&	$\gamma_{\mathrm{us},4}/\mu^{\mathrm{A}}b$\\\hline \rule{0pt}{2.3ex}
		Value	&	$1.95\cdot 10^{-2}$	&	$8.67\cdot 10^{-3}$	&	$3.28\cdot 10^{-3}$	&	$1.14\cdot 10^{-3}$
	\end{tabular}
	\label{tab:Phase_A}
\end{table}
%
%
%
%
%
\section{Illustrative results}
\label{sect:Unreduced_Results}
The purpose of this section is to demonstrate the model's capability to represent dislocation transmission and dislocation induced crack nucleation. The outcome of the competition between these phenomena depends on the material and interface properties. In this context, first results are given to make the reader familiar with the general mechanics of the problem. Two interfaces of different toughness are considered, one that promotes transmission ($k_{\mathrm{pb}}=0.435$) and one that is prone to failure ($k_{\mathrm{pb}}=0.379$). Throughout this section, a phase contrast of $k_{\mathrm{m}}=2$ applies. 
First, the results of a single dislocation interacting with the phase boundary are given to understand the influence of the decohering phase boundary. Subsequently, an 8-dislocation pile-up system is considered to demonstrate the difference in model response with respect to the single dislocation case. Results are compared with a corresponding non-damaging model ($k_{\mathrm{pb}}=\infty$) where displacement and traction continuity are enforced across $\Gamma_{\mathrm{pb}}$. 
\subsection{Single dislocation case}
\label{sect:disl-vs-pb-wo_Rice}
\begin{figure}[htbp]
	\centering
	\includegraphics[width=1.0\linewidth]{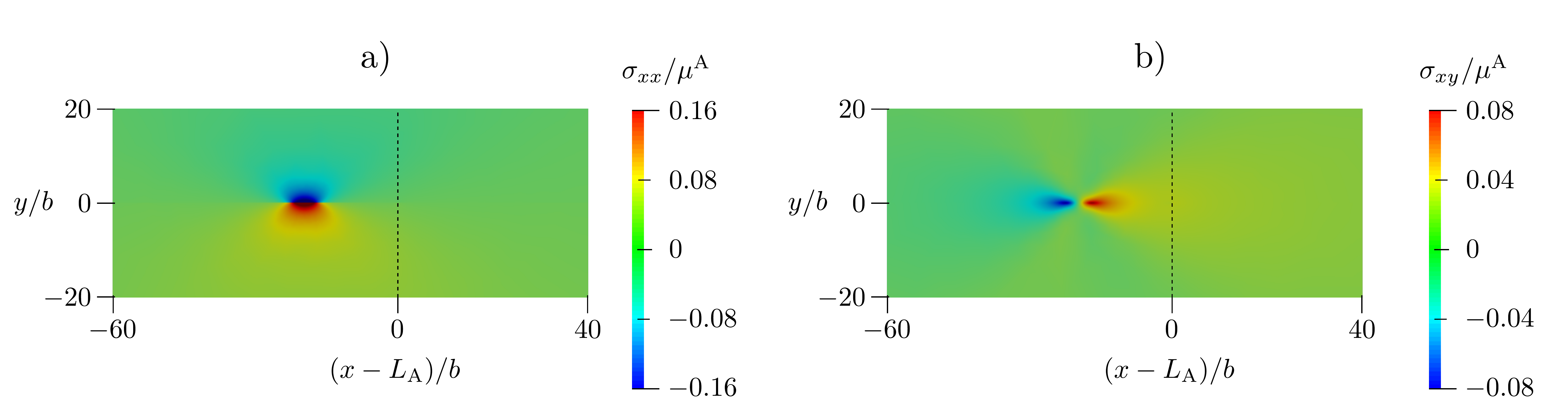}
	\caption{Stress field of a single dislocation interacting with a decohering phase boundary ($k_{\mathrm{pb}}=0.435$) at $\tau=0.0019\mu^{\mathrm{A}}$: (a) $\sigma_{xx}$ and (b) $\sigma_{xy}$. The phase boundary is indicated by the dashed line.}
	\label{fig:1-disl-tau1}
\end{figure}

Consider first a single dislocation under the externally applied shear load $\tau$. While the shear load acts as a driving force on the dislocation towards the phase boundary, a repulsive image force arises from the phase contrast between the two phases, creating a natural source of dislocation obstruction. Equilibrium is attained, for a given level of applied shear, when these two forces are in equilibrium. This is illustrated in Figure \ref{fig:1-disl-tau1} for $\tau=0.0019\,\mu^{\mathrm{A}}$ and $k_{\mathrm{pb}}=0.435$ by the stress fields $\sigma_{xx}$ and $\sigma_{xy}$. At this applied shear load, a dislocation equilibrium position is established at approximately $30\,b$ from the phase boundary.\par

\begin{figure}[htbp]
	\centering
	\includegraphics[width=1.0\linewidth]{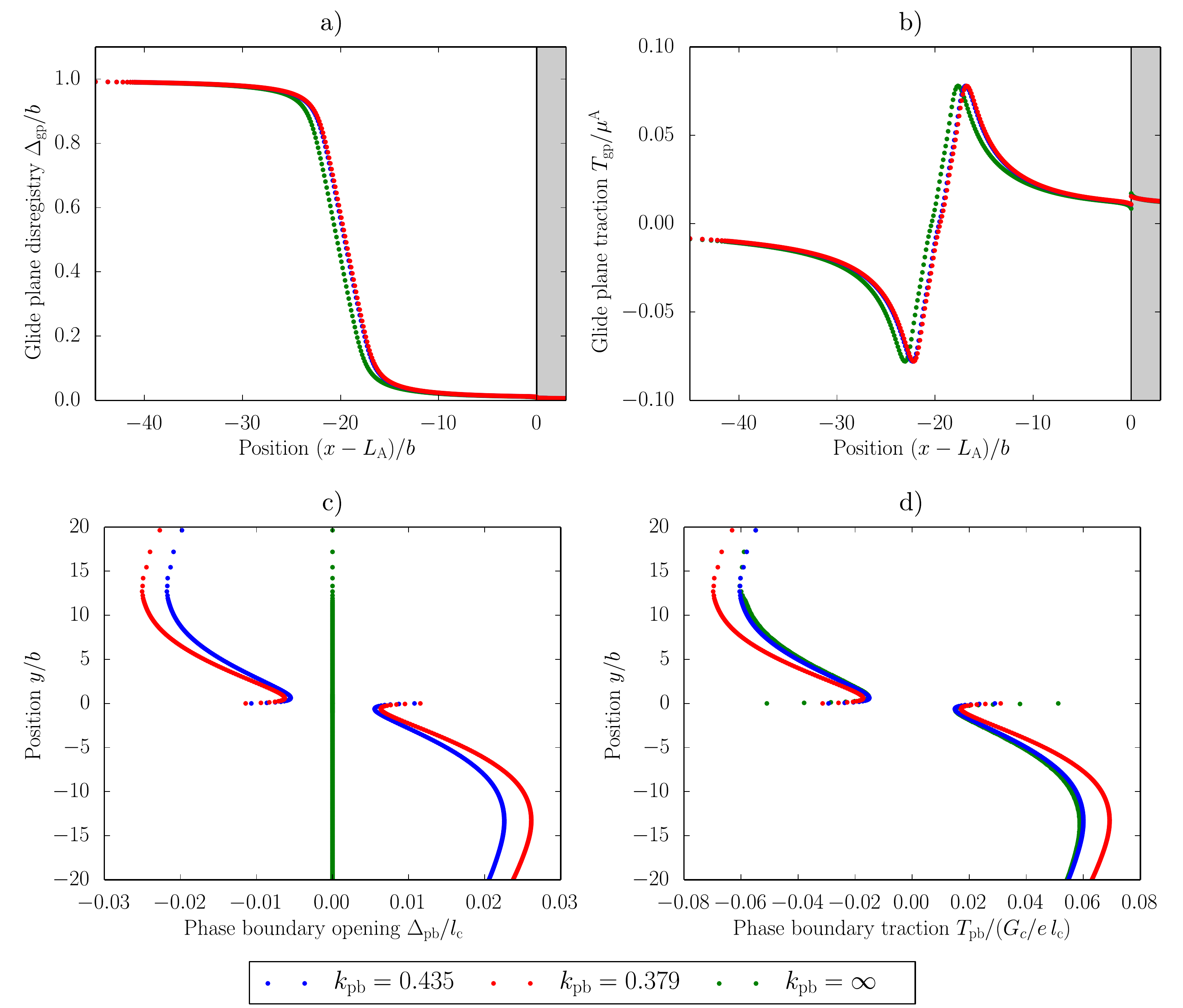}
	\caption{Model response for a single dislocation interacting with a non-damaging ($k_{\mathrm{pb}}=\infty$) and a decohering phase boundary ($k_{\mathrm{pb}}\in\left\{0.379,0.435\right\}$) at $\tau=0.0019\mu^{\mathrm{A}}$: (a) disregistry profiles $\Delta_{\mathrm{gp}}$ and (b) glide plane tractions $T_{\mathrm{gp}}$ along the glide plane. Phase~$\mathrm{B}$ is shaded for clarity. (c) Opening profiles $\Delta_{\mathrm{pb}}$ and (d) phase boundary tractions $T_{\mathrm{pb}}$ along the phase boundary plane.}
	\label{fig:gp-pb-wo_Rice-inc-2}
\end{figure}

For the comparison of the model responses for different toughness factors, the results are displayed in terms of the glide plane response and of the phase boundary response in Figure \ref{fig:gp-pb-wo_Rice-inc-2}. The glide plane behaviour is illustrated by the disregistry profiles $\Delta_{\mathrm{gp}}$ (Figure \ref{fig:gp-pb-wo_Rice-inc-2}a) and the glide plane tractions $T_{\mathrm{gp}}$ (Figure \ref{fig:gp-pb-wo_Rice-inc-2}b) which for $k_{\mathrm{pb}}=0.435$ and $k_{\mathrm{pb}}=0.379$ are nearly overlapping. However, a slight deviation is observed from the profile obtained for the perfectly bonded case ($k_{\mathrm{pb}}=\infty$). The presence of the dislocation is indicated by the drop of the disregistry from $b$ to $0$, which is established by the energy minimisation -- without requiring any additional criteria. The dislocation core is located at the position where $\Delta_{\mathrm{gp}}=b/2$. The related glide plane tractions are also an outcome of the simulation. They remain finite and are zero at the centre of the dislocation. Note that the discontinuity in $T_{\mathrm{gp}}$ at the phase boundary $x=L_{\mathrm{A}}$ originates from the jump in the piece-wise constant material properties across the phase boundary.\par
The response of the phase boundary is demonstrated by the opening profiles $\Delta_{\mathrm{pb}}$ (Figure \ref{fig:gp-pb-wo_Rice-inc-2}c) and the phase boundary tractions $T_{\mathrm{pb}}$ (Figure \ref{fig:gp-pb-wo_Rice-inc-2}d). Here, the relatively large opening and traction gradients around $y=0$ are induced by the interaction of the phase boundary with the glide plane in relation with a marginal (but barely visible) disregistry gradient at $x=L_{\mathrm{A}}$ .\par
The comparison of the different model responses shows a small influence of the phase boundary opening, which is explained as follows. Due to the presence of the dislocation, a stress field is induced which leads to a slight opening of the phase boundary ($y\le 0$) or a slight compression ($y\ge 0$). Hence, the bulk $\Omega_\pm$ relaxes, resulting in a dislocation position slightly closer to the boundary than for $k_{\mathrm{pb}}=\infty$. At this applied shear load, the small difference between $k_{\mathrm{pb}}=0.435$ and $k_{\mathrm{pb}}=0.379$ has only a negligible influence on $\Delta_{\mathrm{gp}}$ and $T_{\mathrm{gp}}$, and hence on the dislocation position.\par
\begin{figure}[htbp]
	\centering
	\includegraphics[width=1.0\linewidth]{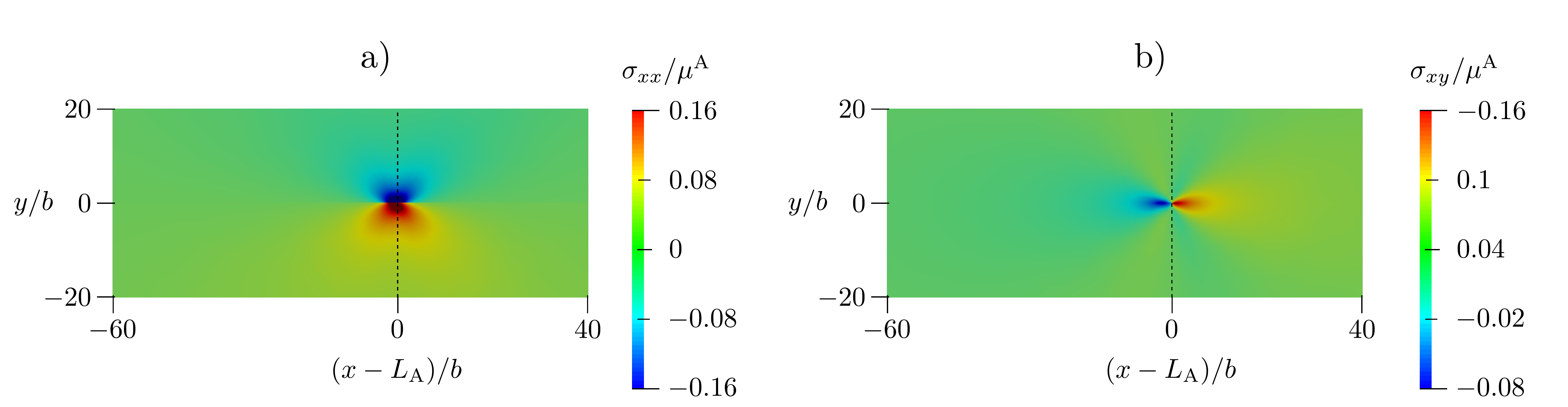}
	\caption{Stress field of a single dislocation interacting with a decohering phase boundary ($k_{\mathrm{pb}}=0.435$) at $\tau=0.04\mu^{\mathrm{A}}$: (a) $\sigma_{xx}$ and (b) $\sigma_{xy}$.}
	\label{fig:1-disl-tau2}
\end{figure}
\begin{figure}[htbp]
	\centering
	\includegraphics[width=1.0\linewidth]{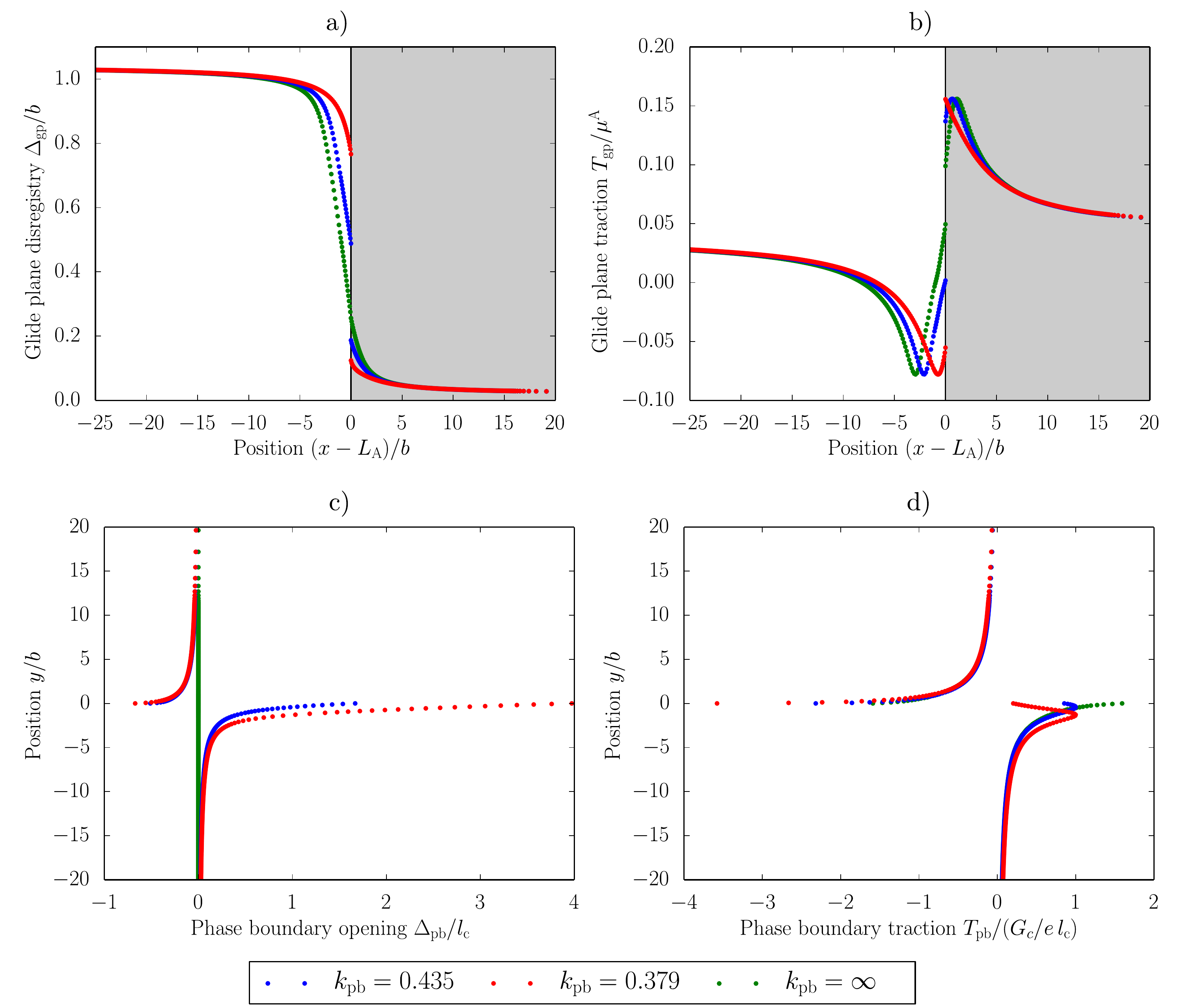}
	\caption{Model response for a single dislocation interacting with a non-damaging ($k_{\mathrm{pb}}=\infty$) and a decohering phase boundary ($k_{\mathrm{pb}}\in\left\{0.379,0.435\right\}$) at $\tau=0.04\mu^{\mathrm{A}}$: (a) disregistry profiles $\Delta_{\mathrm{gp}}$ and (b) glide plane tractions $T_{\mathrm{gp}}$ along the glide plane. (c) Opening profiles $\Delta_{\mathrm{pb}}$ and (d) phase boundary tractions $T_{\mathrm{pb}}$ along the phase boundary plane.}
	\label{fig:gp-pb-wo_Rice-inc-6}
\end{figure}

With an increasing externally applied shear load, the influence of the phase boundary opening becomes more pronounced. To observe this, consider the single dislocation response under an externally applied shear load of $\tau=0.04\,\mu^{\mathrm{A}}$, as illustrated for $k_{\mathrm{pb}}=0.435$ in Figure \ref{fig:1-disl-tau2} in terms of the stress fields $\sigma_{xx}$ and $\sigma_{xy}$.
The specific responses of the glide plane and the phase boundary are shown in Figure \ref{fig:gp-pb-wo_Rice-inc-6}. Like before, the glide plane behaviour is presented by the disregistry profiles $\Delta_{\mathrm{gp}}$ (Figure \ref{fig:gp-pb-wo_Rice-inc-6}a) and the glide plane tractions $T_{\mathrm{gp}}$ (Figure \ref{fig:gp-pb-wo_Rice-inc-6}b), and the phase boundary behaviour by the opening profiles $\Delta_{\mathrm{pb}}$ (Figure \ref{fig:gp-pb-wo_Rice-inc-6}c) and the phase boundary tractions $T_{\mathrm{pb}}$ (Figure \ref{fig:gp-pb-wo_Rice-inc-6}d). \par
Due to the proximity of the dislocation to the phase boundary, the dislocation induced tractions on $\Gamma_{\mathrm{pb}}$ are now higher, which leads to a larger phase boundary opening (note the different horizontal scale in Figure \ref{fig:gp-pb-wo_Rice-inc-6}c-d compared to Figure \ref{fig:gp-pb-wo_Rice-inc-2}c-d). The bulk $\Omega_\pm$ relaxes more, and again the dislocation moves closer to the phase boundary. Naturally, a weaker phase boundary (lower $k_{\mathrm{pb}}$) entails a larger phase boundary opening. Another consequence of the bulk relaxation is a lower dislocation induced net shear stress on the glide plane of Phase~$\mathrm{B}$ (see Figure \ref{fig:gp-pb-wo_Rice-inc-6}b) which leads to the decreased disregistry $\Delta_{\mathrm{gp}}(x^{\mathrm{B}})$, where $x^{\mathrm{B}}=\left\{x\in\Gamma_{\mathrm{gp}}^{\mathrm{B}}|x=L_{\mathrm{A}}\right\}$. Hence, to reach the same net dislocation induced shear stress in Phase~$\mathrm{B}$, as required for transmission, the externally applied shear needs to be increased. While for the non-damaging phase boundary ($k_{\mathrm{pb}}=\infty$) an external transmission stress, i.e. the externally applied shear load at dislocation transmission, of $\tau_{\mathrm{trans}}^{\infty}\approx 0.045\,\mu^{\mathrm{A}}$ is recorded, an increased externally applied shear load of $\tau_{\mathrm{trans}}\approx 1.16\,\tau_{\mathrm{trans}}^{\infty}$ is required for $k_{\mathrm{pb}}=0.435$. With a toughness factor of $k_{\mathrm{pb}}=0.379$, the relaxation is strong enough to inhibit transmission for any externally applied shear load below $\tau=\bar{\tau}$.\par
\subsection{Dislocation pile-up}
\begin{figure}[htbp]
	\centering
	\includegraphics[width=1.0\linewidth]{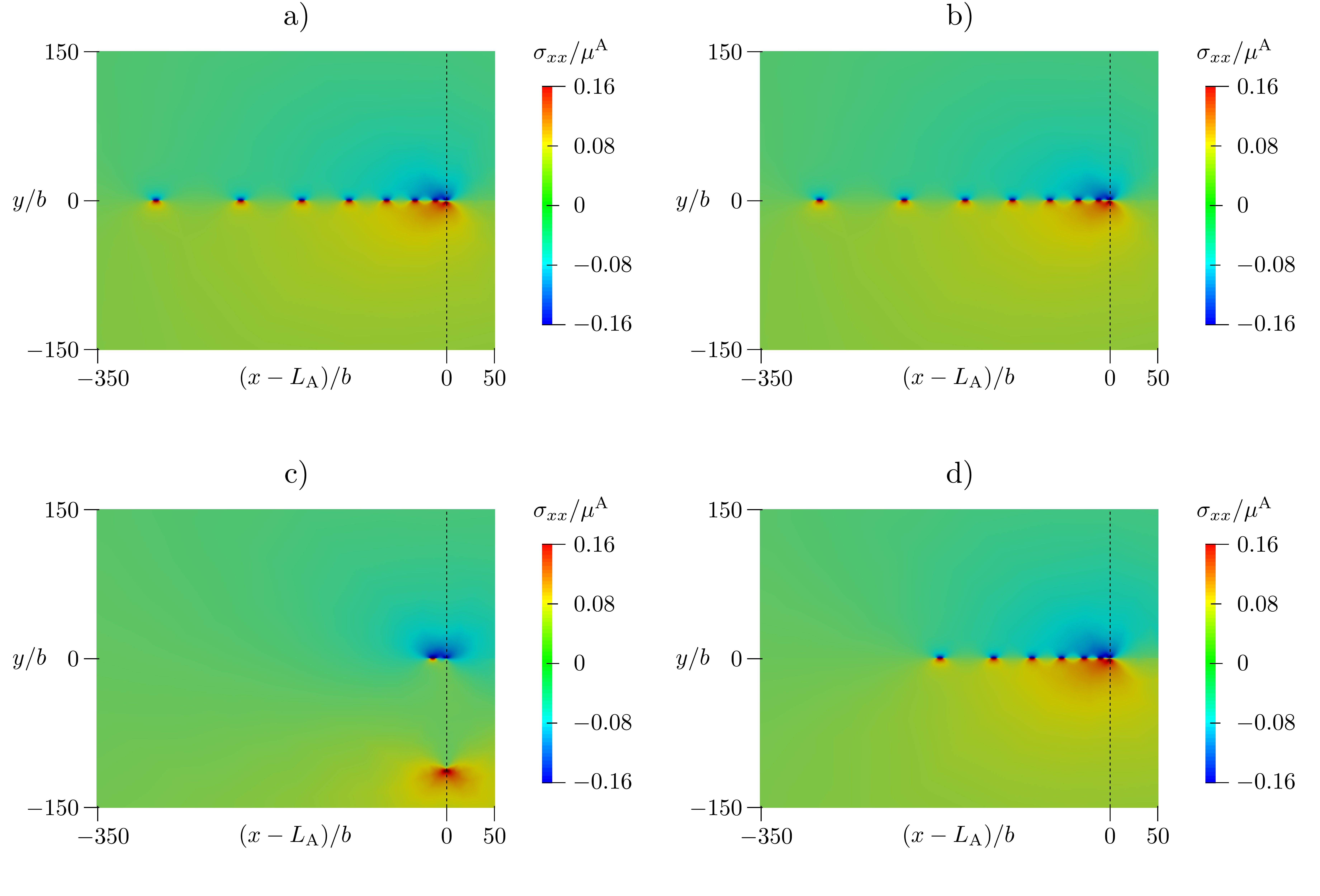}
	\caption{Stress field $\sigma_{xx}$ for an 8 dislocation pile-up system and a decohering phase boundary with the phase contrast $k_{\mathrm{m}}=2$ and toughness factors $k_{\mathrm{pb}}=0.379$ (a,c) and $k_{\mathrm{pb}}= 0.435$ (b,d) under different externally applied shear loads $\tau$: (a-b) at $\tau=0.0118\,\mu^{\mathrm{A}}$, (c) after crack nucleation at $\tau=0.0182\,\mu^{\mathrm{A}}$ and (d) after transmission at $\tau=0.0181\,\mu^{\mathrm{A}}$.}
	\label{fig:pu-evolutions}
\end{figure}

To demonstrate the capability of the PN-CZ model to simulate the competition between dislocation transmission and phase boundary decohesion as a function of the material properties (including the cohesive properties of the phase boundary), an 8-dislocation pile-up system is now considered. The same material properties as for the single dislocation case apply, with $k_{\mathrm{pb}} \in \left\{0.379,0.435\right\}$. An increasing external shear load $\tau$ is applied, until eventually either a dislocation is transmitted or a crack is nucleated, as illustrated in Figure \ref{fig:pu-evolutions} in terms of the stress field $\sigma_{xx}$ before the event (Figure \ref{fig:pu-evolutions}a-b) and after the event (Figure \ref{fig:pu-evolutions}c-d). Note that in Figure \ref{fig:pu-evolutions}d the first dislocation has been transmitted and hence is no longer visible in the plotted window. \par
The transmission of a dislocation is recorded by its presence in Phase~$\mathrm{B}$, i.e. $\Delta_{\mathrm{gp}}^{\mathrm{B}}>b/2$. A crack is assumed to be nucleated as soon as two dislocations are absorbed by the phase boundary, which corresponds to an opening $\Delta_{\mathrm{pb}}(y=0^-)>3b/2$. In the reference case ($k_{\mathrm{pb}} = \infty$), dislocation transmission occurs at $\tau_{\mathrm{trans}}^{\infty}\approx 0.012\,\mu^{\mathrm{A}}$.\par
\begin{figure}[htbp]
	\centering
	\includegraphics[width=1.0\linewidth]{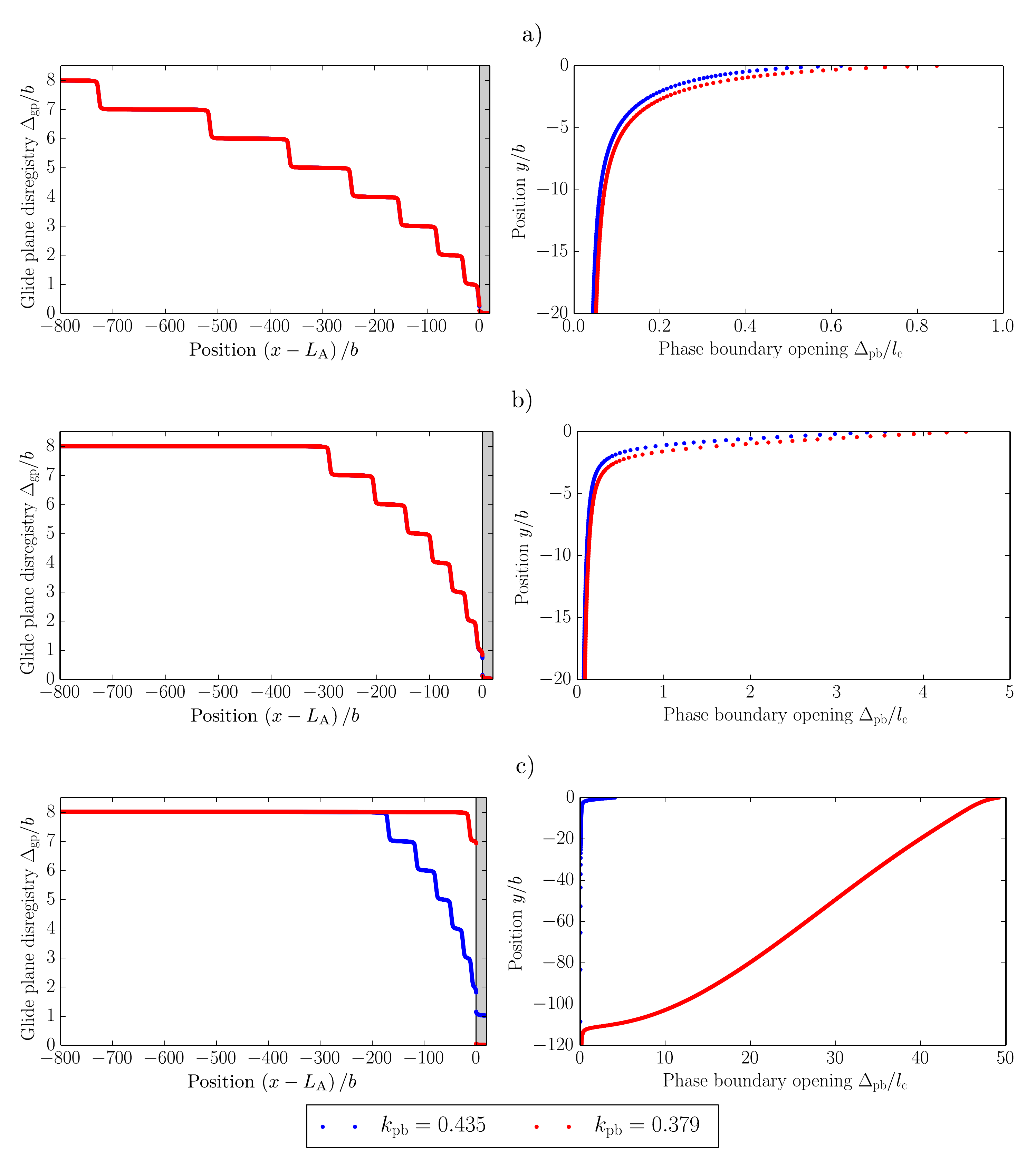}
	\caption{Disregistry and opening profiles $\Delta_{\mathrm{gp}}$ and $\Delta_{\mathrm{pb}}$ for an 8 dislocation pile-up system and a decohering phase boundary with the phase contrast $k_{\mathrm{m}}=2$ and toughness factors $k_{\mathrm{pb}}\in\left\{0.379, 0.435\right\}$ under different externally applied shear loads $\tau$: (a) at $\tau=0.0054\,\mu^{\mathrm{A}}$, (b) at $\tau=0.0118\,\mu^{\mathrm{A}}$ and (c) after crack nucleation at $\tau=0.0182\,\mu^{\mathrm{A}}$ for $k_{\mathrm{pb}}=0.379$ and after transmission at $\tau=0.0181\,\mu^{\mathrm{A}}$ for $k_{\mathrm{pb}}=0.435$. Note the different scale for the opening.}
	\label{fig:Disregistry-8pu}
\end{figure}

The evolution of the glide plane and phase boundary responses for the 8-dislocation pile-up, including dislocation transmission and crack nucleation, are plotted in Figure \ref{fig:Disregistry-8pu} in terms of the disregistry and opening profiles at different externally applied shear loads $\tau$. Similar to the single dislocation case, the position of dislocation $j$ is where $\Delta_{\mathrm{gp}}=(2j - 1)b/2$.\par
For both phase boundary toughnesses, the dislocation pile-up evolves similarly before either event (transmission or decohesion) is triggered. Only the opening behaviour shows a small mismatch, due to the different phase boundary toughness. Ultimately, under sufficient load on the pile-up, the model responses deviate, exhibiting either dislocation transmission ($k_{\mathrm{pb}}=0.435$) or phase boundary decohesion ($k_{\mathrm{pb}}=0.379$), at $\tau_{\mathrm{trans}}\approx 1.51\,\tau_{\mathrm{trans}}^{\infty}$ and $\tau_{\mathrm{dec}}\approx 1.52\,\tau_{\mathrm{trans}}^{\infty}$, respectively. In the case of crack nucleation, an instantaneous propagation occurs until 7 dislocations are absorbed.
These preliminary results show that the PN-CZ model is fully capable of capturing the competition between dislocation transmission and phase boundary decohesion.
%
%
%
%
\section{Reduced interfacial potentials}
\label{sect:Reduced_Potential}
\subsection{Methodology}
\begin{figure}[htbp]
	\centering
	\includegraphics[width=1.\linewidth]{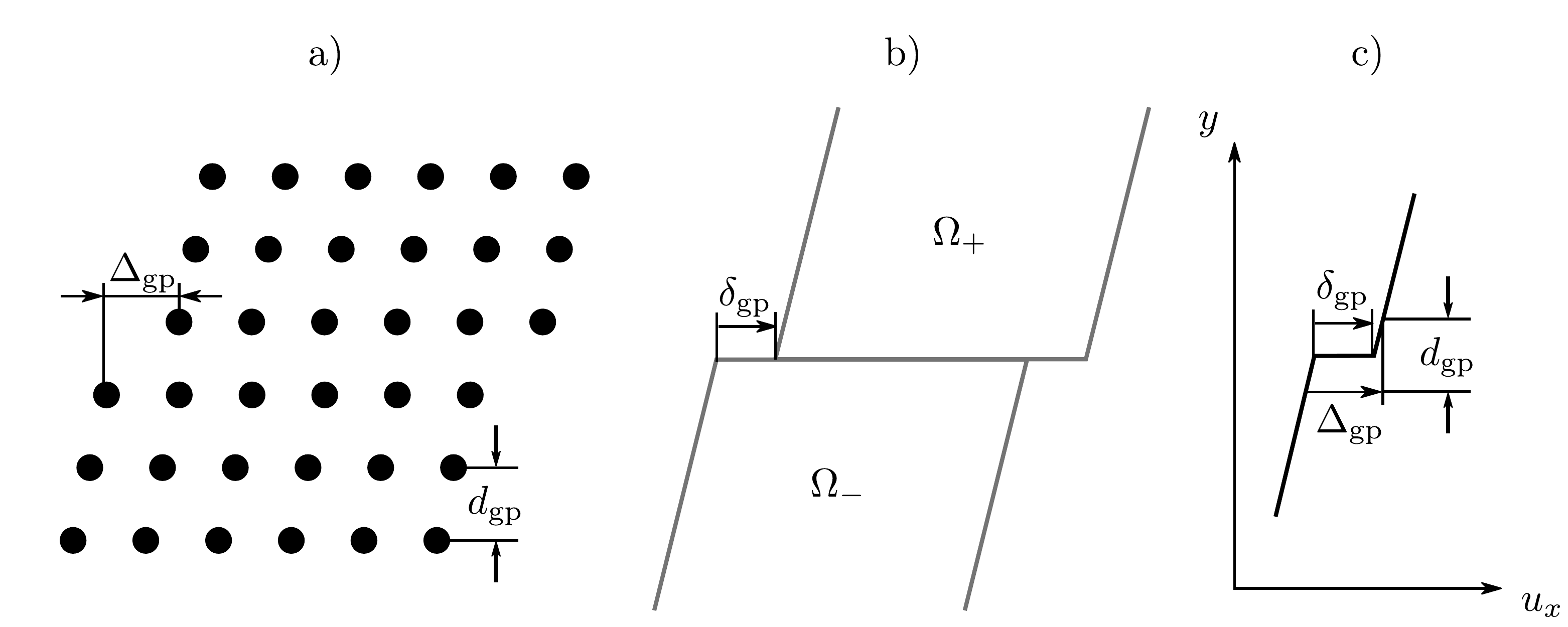}
	\caption{(a) Unreduced glide plane disregistry $\Delta_{\mathrm{gp}}$ in a square lattice, (b) reduced glide plane disregistry $\delta_{\mathrm{gp}}$ in the PN-CZ model and (c) the physical relation between $\Delta_{\mathrm{gp}}$ and $\delta_{\mathrm{gp}}$.}
	\label{fig:Rice_gp}
\end{figure}

In Section \ref{sect:Unreduced_Results} it has been shown that the PN-CZ model is capable of capturing the competition between dislocation transmission and phase boundary decohesion. Atomistically calculated material properties have been adopted to describe the bulk ($\Omega_\pm^i$) behaviour, as well as the behaviour of the glide plane ($\Gamma_{\mathrm{gp}}^i$) and the phase boundary ($\Gamma_{\mathrm{pb}}$), both modelled as zero-thickness interfaces. The atomistic potentials for the glide plane and the phase boundary, however, correspond to a misfit energy which is induced by the rigid shift $\Delta_{\mathrm{gp}}$ or $\Delta_{\mathrm{pb}}$ between two bulks of atoms adjacent to the interface ($\Gamma_{\mathrm{gp}}$ or $\Gamma_{\mathrm{pb}}$), as illustrated for the glide plane in Figure \ref{fig:Rice_gp}a. Thus, by assigning these potentials to the zero-thickness interfaces, an error has been introduced due to the inclusion of the (linear) elastic response of the thin layer of thickness $d_{\mathrm{gp}}$ (for $\Gamma_{\mathrm{gp}}$) or $d_{\mathrm{pb}}$ (for $\Gamma_{\mathrm{pb}}$) into the (zero-thickness) interface model. To rectify this physical inconsistency, Rice \cite{Rice1992} and later Sun et. al \cite{Sun1993} proposed the exclusion of this linear elastic response from the atomistically calculated potentials to obtain the corresponding non-linear potentials of the zero-thickness interfaces. In this context, by subtracting the linear elastic displacement from unreduced disregistry $\Delta_{\mathrm{gp}}$ and opening $\Delta_{\mathrm{pb}}$, the reduced disregistry $\delta_{\mathrm{gp}}$ and opening $\delta_{\mathrm{pb}}$ of the zero-thickness interface are obtained, as illustrated for the glide plane in Figure \ref{fig:Rice_gp}b and \ref{fig:Rice_gp}c. \par
Let the interface potentials $\psi_{\mathrm{gp}}$ and $\psi_{\mathrm{pb}}$ as obtained from atomistics, which are considered as given, be comprised of an elastic contribution $\psi_{\mathrm{gp,e}}$ and $\psi_{\mathrm{pb,e}}$ (intrinsic to the half-bands above and below the zero-thickness interface) and the reduced potentials $\psi_{\mathrm{gp}}^*$ and $\psi_{\mathrm{pb}}^*$ of the connecting zero-thickness interface:
\begin{align}
	\label{eq:reduced_gp_potential}
	\psi_{\mathrm{gp}}(\Delta_{\mathrm{gp}}) =& \psi_{\mathrm{gp,e}}(\Delta_{\mathrm{gp}}, \delta_{\mathrm{gp}}) + \psi_{\mathrm{gp}}^*(\delta_{\mathrm{gp}})\\
	\label{eq:reduced_pb_potential}
	\psi_{\mathrm{pb}}(\Delta_{\mathrm{pb}}) =& \psi_{\mathrm{pb,e}}(\Delta_{\mathrm{pb}}, \delta_{\mathrm{pb}}) + \psi_{\mathrm{pb}}^*(\delta_{\mathrm{pb}})
\end{align}
The elastic contribution of the band is defined for a linear elastic solid as
\begin{align}
	\label{eq:gp-el-energy}
	\psi_{\mathrm{gp,e}}= & \frac{1}{2}\frac{\mu_{\mathrm{gp}}}{d_{\mathrm{gp}}}\left(\Delta_{\mathrm{gp}}-\delta_{\mathrm{gp}}\right)^2\\
	\label{eq:pb-el-energy}
	\psi_{\mathrm{pb,e}}= & \frac{1}{2}\frac{c_{\mathrm{pb}}}{d_{\mathrm{pb}}}\left(\Delta_{\mathrm{pb}}-\delta_{\mathrm{pb}}\right)^2
\end{align}
where $\mu_{\mathrm{gp}}$ and $c_{\mathrm{pb}}$ are the shear modulus and uniaxial strain modulus, respectively. For infinitesimal disregistries $\Delta_{\mathrm{gp}}$ and openings $\Delta_{\mathrm{pb}}$, the response of the potentials $\psi_{\mathrm{gp}}$ and $\psi_{\mathrm{pb}}$ can be considered as linear elastic only. Requiring this limit behaviour implies for $\psi_{\mathrm{gp}}$, $\psi_{\mathrm{pb}}$ and $\psi_{\mathrm{gp,e}}$, $\psi_{\mathrm{pb,e}}$ 
\begin{align}
	\label{eq:gp-stiffness}
	M_{\mathrm{gp},0}:=\left.\frac{\mathrm{d}^2\psi_{\mathrm{gp}}}{\mathrm{d}\Delta_{\mathrm{gp}}^2}\right|_{\Delta_{\mathrm{gp}}=0}=&\frac{\mathrm{d}^2\psi_{\mathrm{gp,e}}}{\mathrm{d}\Delta_{\mathrm{gp}}^2}=\frac{\mu_{\mathrm{gp}}}{d_{\mathrm{gp}}}\\
	\label{eq:pb-stiffness}
	M_{\mathrm{pb},0}:=\left.\frac{\mathrm{d}^2\psi_{\mathrm{pb}}}{\mathrm{d}\Delta_{\mathrm{pb}}^2}\right|_{\Delta_{\mathrm{pb}}=0}=&\frac{\mathrm{d}^2\psi_{\mathrm{gp,e}}}{\mathrm{d}\Delta_{\mathrm{gp}}^2}=\frac{c_{\mathrm{pb}}}{d_{\mathrm{pb}}}
\end{align}
Note that in relation with the rigid shift of the two bulks of atoms with respect to each other, $\mu_{\mathrm{gp}}$ and $c_{\mathrm{pb}}$ do not exactly correspond to the homogeneous bulk properties $\mu$ and $c$. The reduced potentials for the zero-thickness interfaces $\Gamma_{\mathrm{gp}}$ and $\Gamma_{\mathrm{pb}}$ follow from Eq. \eqref{eq:reduced_gp_potential}-\eqref{eq:pb-stiffness}:
\begin{align}
	\label{eq:gp-red_energy}
	\psi_{\mathrm{gp}}^*(\delta_{\mathrm{gp}})= & \psi_{\mathrm{gp}}(\Delta_{\mathrm{gp}})-\frac{1}{2}M_{\mathrm{gp},0}\left(\Delta_{\mathrm{gp}}-\delta_{\mathrm{gp}}\right)^2\\
	\label{eq:pb-red_energy}
	\psi_{\mathrm{pb}}^*(\delta_{\mathrm{pb}})= & \psi_{\mathrm{pb}}(\Delta_{\mathrm{pb}})-\frac{1}{2}M_{\mathrm{pb},0}\left(\Delta_{\mathrm{pb}}-\delta_{\mathrm{pb}}\right)^2
\end{align}
The total free energy of Eq. \eqref{eq:internal-energy} is modified accordingly with the reduced potentials. The reduced disregistry $\delta_{\mathrm{gp}}$ and opening and $\delta_{\mathrm{pb}}$ replace the unreduced counterparts as primary dependent variables and are defined as the relative displacements 
\begin{align}
	\label{eq:gp-red-disregistry}
	\delta_{\mathrm{gp}} = & [[\vec{u}]] \cdot \vec{e}_x, && \vec{x}\in\Gamma_{\mathrm{gp}}\\
	\label{eq:pb-red-disregistry}
	\delta_{\mathrm{pb}} = & [[\vec{u}]] \cdot \vec{e}_x, && \vec{x}\in\Gamma_{\mathrm{pb}}
\end{align}
Yet, the unreduced disregistry $\Delta_{\mathrm{gp}}$ and opening and $\Delta_{\mathrm{pb}}$ are required to calculate the reduced potentials. The link between the reduced and unreduced disregistries and openings is established through the differentiation of Eq. \eqref{eq:gp-red_energy} and \eqref{eq:pb-red_energy} with respect to $\Delta_{\mathrm{gp}}$ and $\Delta_{\mathrm{pb}}$, respectively, and reads
\begin{align}
	\label{eq:gp-nl-disregistry}
	\delta_{\mathrm{gp}}=& \Delta_{\mathrm{gp}}-\frac{1}{M_{\mathrm{gp},0}}T_{\mathrm{gp}}(\Delta_{\mathrm{gp}})\\
	\label{eq:pb-nl-disregistry}
	\delta_{\mathrm{pb}}=& \Delta_{\mathrm{pb}}-\frac{1}{M_{\mathrm{pb},0}}T_{\mathrm{pb}}(\Delta_{\mathrm{pb}})
\end{align}
The unreduced disregistry $\Delta_{\mathrm{gp}}$ and opening $\Delta_{\mathrm{pb}}$ are obtained by solving these non-linear equations iteratively for the given reduced disregistry $\delta_{\mathrm{gp}}$ and opening $\delta_{\mathrm{pb}}$.\par
\begin{figure}[htbp]
	\centering
	\includegraphics[width=1.0\linewidth]{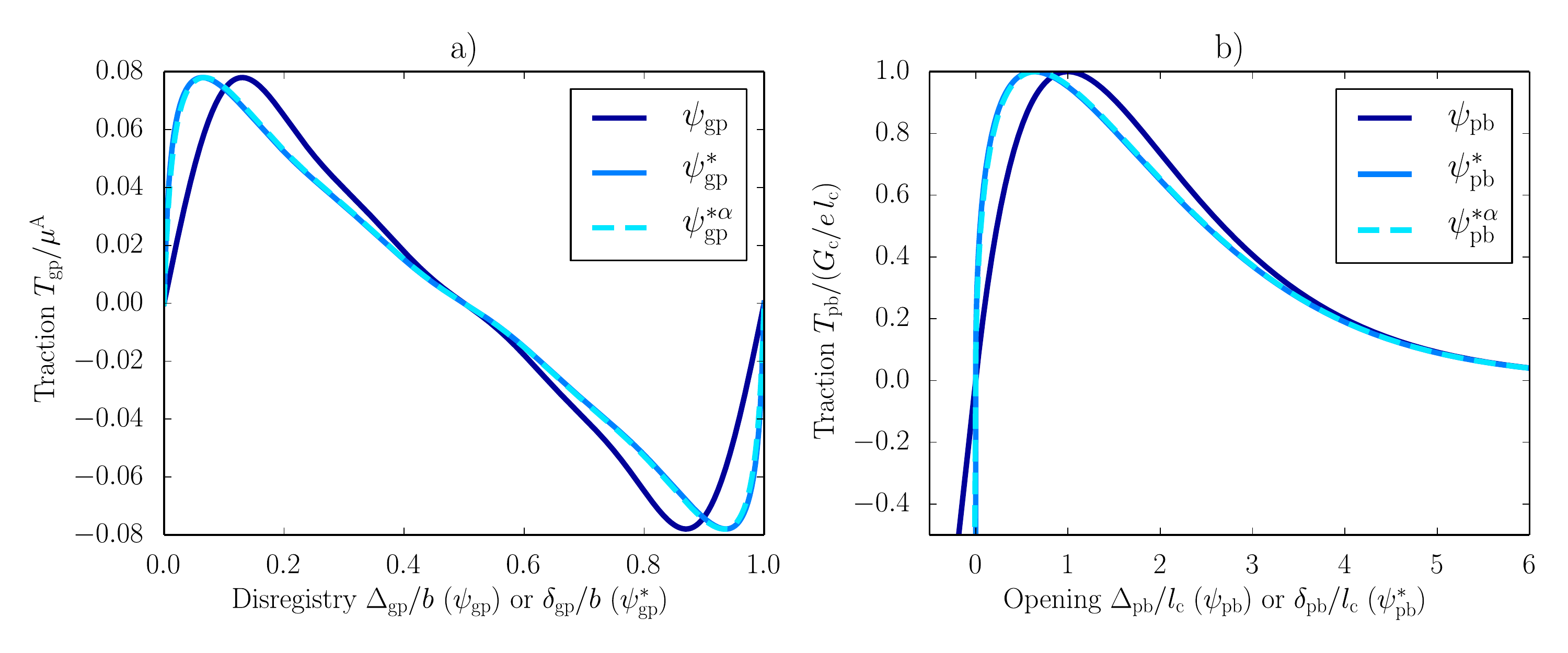}
	\caption{(a) Glide plane traction $T_{\mathrm{gp}}(\Delta_{\mathrm{gp}})$ and $T_{\mathrm{gp}}^*(\delta_{\mathrm{gp}})=T_{\mathrm{gp}}(\Delta_{\mathrm{gp}}(\delta_{\mathrm{gp}}))$ and (b) phase boundary traction profiles $T_{\mathrm{pb}}(\Delta_{\mathrm{pb}})$ and $T_{\mathrm{pb}}^*(\delta_{\mathrm{pb}})=T_{\mathrm{pb}}(\Delta_{\mathrm{pb}}(\delta_{\mathrm{pb}}))$. $T_{\mathrm{gp}}$ and $T_{\mathrm{pb}}$ correspond to the conventional definition of the glide plane and cohesive zone, whereas $T_{\mathrm{gp}}^*$ and $T_{\mathrm{pb}}^*$ refer to their reduced counterparts from which the linear elastic response has been eliminated. The regularised tractions (Eq. \eqref{eq:gp-red_energy-reg}-\eqref{eq:pb-nl-disregistry-reg}) with $\alpha=0.95$ are labelled as $\psi_{\mathrm{gp}}^{*\alpha}$ and $\psi_{\mathrm{pb}}^{*\alpha}$. }
	\label{fig:Traction-profiles}
\end{figure}

As a result of this linear elastic reduction, the physical consistency of the zero-thickness character of the interfaces of the PN-CZ model is recovered, i.e. the initial compliance for $\delta_{\mathrm{gp}}=i\,b$ ($i=1,2,\dots$) and $\delta_{\mathrm{pb}}=0$ is zero. This is illustrated in Figure \ref{fig:Traction-profiles}a by the glide plane tractions $T_{\mathrm{gp}}$ and $T_{\mathrm{gp}}^*$ as a function of the disregistries $\Delta_{\mathrm{gp}}$ and $\delta_{\mathrm{gp}}$, respectively, and in Figure \ref{fig:Traction-profiles}b by the phase boundary traction $T_{\mathrm{pb}}$, $T_{\mathrm{pb}}^*$ as a function of the openings $\Delta_{\mathrm{pb}}$ and $\delta_{\mathrm{pb}}$, respectively.\par
A complication in the numerical implementation is that the linear elastic reduction leads to zero compliance (and infinite stiffness) at $\delta_{\mathrm{gp}}=i\,b$ ($i=1,2,\dots$) and $\delta_{\mathrm{pb}}=0$, resulting in an ill-condition Hessian. To facilitate the numerical solution, it is therefore regularised. The reduced potentials and the link between the reduced and unreduced disregistry and opening  are modified to
\begin{align}
	\label{eq:gp-red_energy-reg}
	\psi_{\mathrm{gp}}^*(\delta_{\mathrm{gp}})= & \psi_{\mathrm{gp}}(\Delta_{\mathrm{gp}})-\frac{1}{2\alpha_{\mathrm{r}}}M_{\mathrm{gp},0}\left(\Delta_{\mathrm{gp}}-\delta_{\mathrm{gp}}\right)^2\\
	\psi_{\mathrm{pb}}^*(\delta_{\mathrm{pb}})= & \psi_{\mathrm{pb}}(\Delta_{\mathrm{pb}})-\frac{1}{2\alpha_{\mathrm{r}}}M_{\mathrm{pb},0}\left(\Delta_{\mathrm{pb}}-\delta_{\mathrm{pb}}\right)^2
\end{align}
and
\begin{align}
	\label{eq:gp-nl-disregistry-reg}
	\delta_{\mathrm{gp}}=& \Delta_{\mathrm{gp}}-\frac{\alpha_{\mathrm{r}}}{M_{\mathrm{gp},0}}T_{\mathrm{gp}}(\Delta_{\mathrm{gp}})\\
	\label{eq:pb-nl-disregistry-reg}
	\delta_{\mathrm{pb}}=& \Delta_{\mathrm{pb}}-\frac{\alpha_{\mathrm{r}}}{M_{\mathrm{pb},0}}T_{\mathrm{pb}}(\Delta_{\mathrm{pb}})
\end{align}
with the regularisation factor $\alpha_{\mathrm{r}}$. In this paper, $\alpha_{\mathrm{r}}=0.95$ is employed, which leads to a traction response which is practically identical to that of the ideal case $\alpha_{\mathrm{r}}=1$, as observed in Figure \ref{fig:Traction-profiles}, but which is numerically more benign.
%
%
\subsection{Influence of the linear elastic reduction}
%
%
\subsubsection{Single dislocation}
\begin{figure}[htbp]
	\centering
	\includegraphics[width=1.0\linewidth]{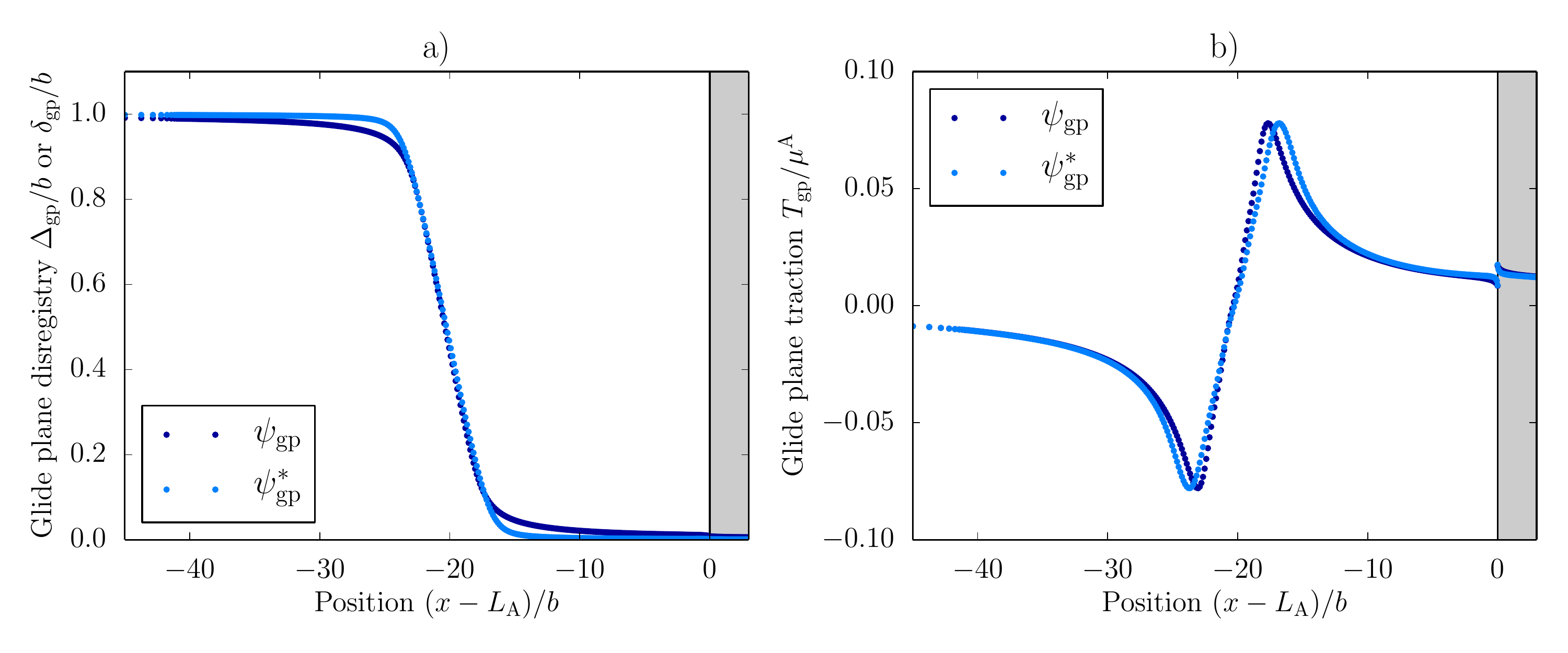}
	\caption{Model response for a single dislocation interacting with a non-damaging phase boundary ($k_{\mathrm{pb}}=\infty$) at $\tau=0.0019\mu^{\mathrm{A}}$: (a) disregistry profile $\Delta_{\mathrm{gp}}$ for the unreduced ($\psi_{\mathrm{gp}}$) and $\delta_{\mathrm{gp}}$ for the reduced potential ($\psi_{\mathrm{gp}}^*$) and (b) the glide plane tractions $T_{\mathrm{gp}}=T_{\mathrm{gp}}^*$.}
	\label{fig:Disregistry-Traction_all-tied-inc-2}
\end{figure}

\begin{figure}[htbp]
	\centering
	\includegraphics[width=1.0\linewidth]{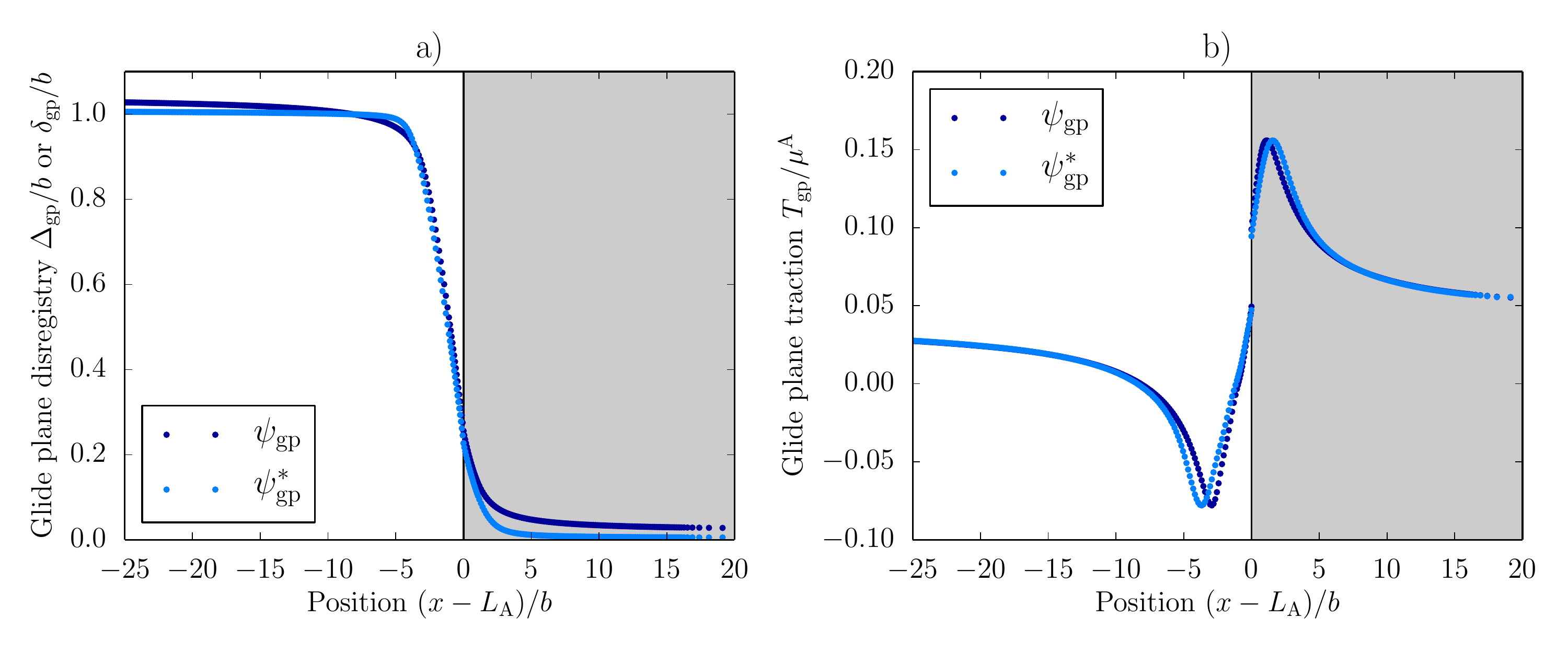}
	\caption{Model response for a single dislocation interacting with a non-damaging phase boundary ($k_{\mathrm{pb}}=\infty$) at $\tau=0.04\mu^{\mathrm{A}}$: (a) disregistry profile $\Delta_{\mathrm{gp}}$ for the unreduced ($\psi_{\mathrm{gp}}$) and $\delta_{\mathrm{gp}}$ for the reduced potential ($\psi_{\mathrm{gp}}^*$) and (b) the glide plane tractions $T_{\mathrm{gp}}=T_{\mathrm{gp}}^*$.}
	\label{fig:Disregistry-Traction_all-tied-inc-6}
\end{figure}

\begin{figure}[htbp]
	\centering
	\includegraphics[width=1.0\linewidth]{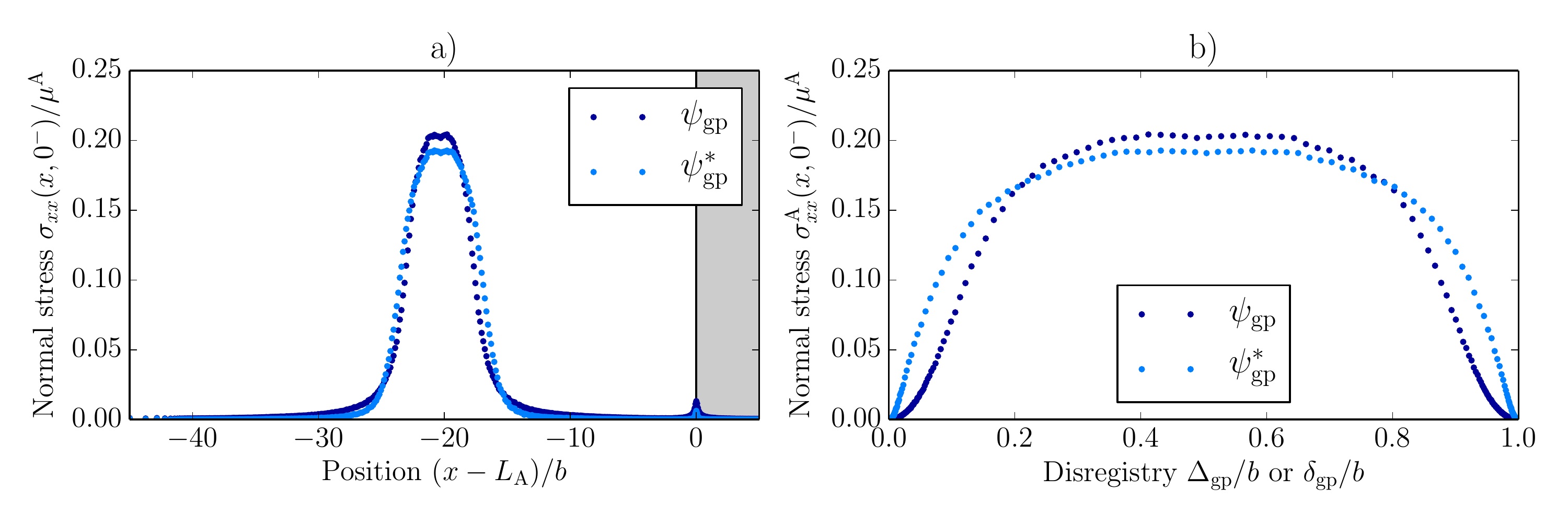}
	\caption{Normal stress distribution $\sigma_{xx}(x,y=0^-)$ for the unreduced ($\psi_{\mathrm{gp}}$) and the reduced potential ($\psi_{\mathrm{gp}}^*$) at $\tau=0.0019\,\mu^{\mathrm{A}}$: (a) $\sigma_{xx}$ as a function of the position $x$ and (b) $\sigma_{xx}$ as a function of $\Delta_{\mathrm{gp}}$, $\delta_{\mathrm{gp}}$ in Phase~$\mathrm{A}$.}
	\label{fig:normal-stress-tied-vs-pos_disreg}
\end{figure}

\paragraph{Non-damaging phase boundary}
To asses the influence of the linear elastic reduction of the potential, consider first the case of a single dislocation approaching a non-damaging phase boundary ($k_{\mathrm{pb}}=\infty$). The phase contrast is set to $k_{\mathrm{m}}=2$. Results show a negligible influence of the potential reduction on the dislocation position and on the external transmission stress, which equals $\tau_{\mathrm{trans}}=0.0451\mu^{\mathrm{A}}$ without and $\tau_{\mathrm{trans}}^*=0.0454\mu^{\mathrm{A}}$ with the reduction applied. Only minor differences can be observed in the disregistries $\Delta_{\mathrm{gp}}$, $\delta_{\mathrm{gp}}$ and the tractions $T_{\mathrm{gp}}(\Delta_{\mathrm{gp}})$, $T_{\mathrm{gp}}^*(\delta_{\mathrm{gp}})=T_{\mathrm{gp}}(\Delta_{\mathrm{gp}}(\delta_{\mathrm{gp}}))$. This is illustrated in Figures \ref{fig:Disregistry-Traction_all-tied-inc-2} and \ref{fig:Disregistry-Traction_all-tied-inc-6} for externally applied shear loads of $\tau=0.0019\,\mu^{\mathrm{A}}$ and $\tau=0.04\,\mu^{\mathrm{A}}$, respectively. The most obvious difference between $\Delta_{\mathrm{gp}}$ and $\delta_{\mathrm{gp}}$ is the vertical offset between both curves (e.g. in Figure \ref{fig:Disregistry-Traction_all-tied-inc-6}a), which is related to the artificial compliance of $\psi_{\mathrm{gp}}$ around $\Delta_{\mathrm{gp}}=i\,b$ and increases with the externally applied shear load $\tau$. In addition, for $\psi_{\mathrm{gp}}^*$ the disregistry profile levels out faster away from the dislocation core due to the difference in compliance. This has a direct influence on the stress distribution, as demonstrated by the shear tractions $T_{\mathrm{gp}}(x)=\sigma_{xy}(x,y=0)$ in Figures \ref{fig:Disregistry-Traction_all-tied-inc-2} and \ref{fig:Disregistry-Traction_all-tied-inc-6}, and by the normal stress $\sigma_{xx}(x,y=0^-)$ in Figure \ref{fig:normal-stress-tied-vs-pos_disreg}. A widening of the stress profile due to the reduction becomes apparent. Furthermore, the peak normal stress $\sigma_{xx}$ along the glide plane slightly decreases, whereas a slightly higher stress is observed for small deviations from $\delta_{\mathrm{gp}}=i\,b$, reflecting the increased gradient for these disregistries.\par
\paragraph{Damaging phase boundary}
Eliminating the initial compliance of the phase boundary in the conventional, unreduced model, the potential reduction may have a significant influence on the model response. This is demonstrated for the single dislocation case with a phase contrast of $k_{\mathrm{m}}=2$ and toughness factors of $k_{\mathrm{pb}}=k_{\mathrm{pb}}^*\in\left\{0.379, 0.435\right\}$.\par
\begin{figure}[htbp]
	\centering
	\includegraphics[width=1.0\linewidth]{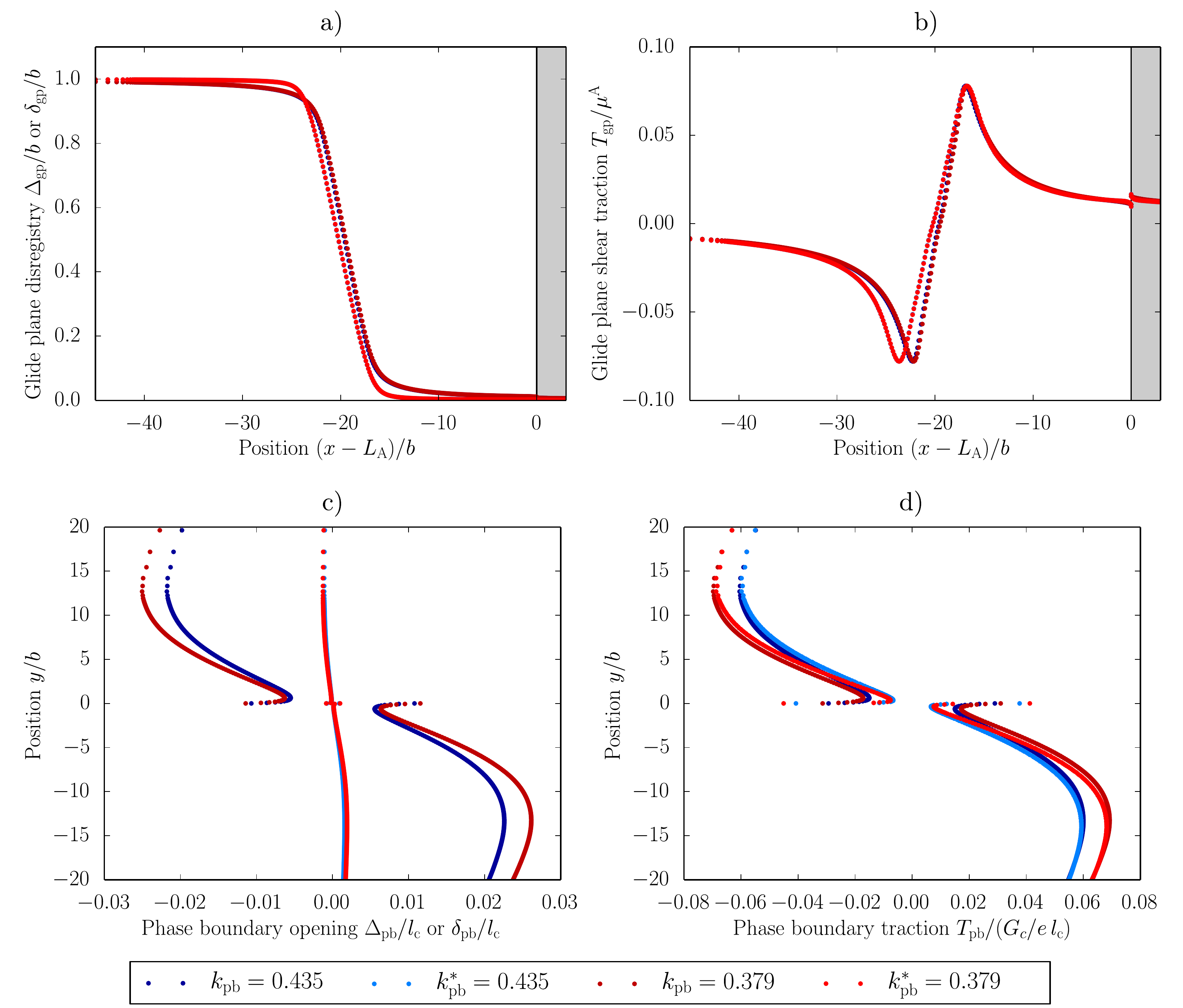}
	\caption{Model response for a single dislocation interacting with a decohering phase boundary with toughness factors $k_{\mathrm{pb}}=k_{\mathrm{pb}}^*\in\left\{0.379,0.435\right\}$ at $\tau=0.0019\mu^{\mathrm{A}}$: (a) disregistry profile $\Delta_{\mathrm{gp}}$ for the unreduced ($\psi_{\mathrm{gp}}$, $\psi_{\mathrm{pb}}$) and $\delta_{\mathrm{gp}}$ for the reduced potentials ($\psi_{\mathrm{gp}}^*$, $\psi_{\mathrm{pb}}^*$) and (b) glide plane traction $T_{\mathrm{gp}}=T_{\mathrm{gp}}^*$. (c) Opening profile $\Delta_{\mathrm{pb}}$ for the unreduced ($\psi_{\mathrm{gp}}$, $\psi_{\mathrm{pb}}$) and $\delta_{\mathrm{pb}}$ for the reduced potentials ($\psi_{\mathrm{gp}}^*$, $\psi_{\mathrm{pb}}^*$) and (d) phase boundary traction $T_{\mathrm{pb}}=T_{\mathrm{pb}}^*$.}
	\label{fig:gp-pb-Rice_comp-inc-2}
\end{figure}

For a dislocation still relatively far from the phase boundary, at $\tau=0.0019\,\mu^{\mathrm{A}}$, minor differences in the dislocation behaviour already arise, as shown in Figure \ref{fig:gp-pb-Rice_comp-inc-2}a-b by the disregistry profiles and the glide plane tractions and in Figure \ref{fig:gp-pb-Rice_comp-inc-2}c-d by the opening profiles and the phase boundary tractions. Note that in Figure \ref{fig:gp-pb-Rice_comp-inc-2}a-b the unreduced and the reduced model responses are independent of the toughness factor. The difference in dislocation position arises from the significantly lower phase boundary compliance of the reduced model around $\delta_{\mathrm{pb}}=0$, invoking only negligible opening and relaxation of the bulk $\Omega^i_\pm$, as opposed to the unreduced potentials. Naturally, the lower bulk relaxation for $k_{\mathrm{pb}}^*$ leads to dislocation positions slightly more distant to the phase boundary and hence a minor decrease in the tractions $T_{\mathrm{pb}}$.
\par
\begin{figure}[htbp]
	\centering
	\includegraphics[width=1.0\linewidth]{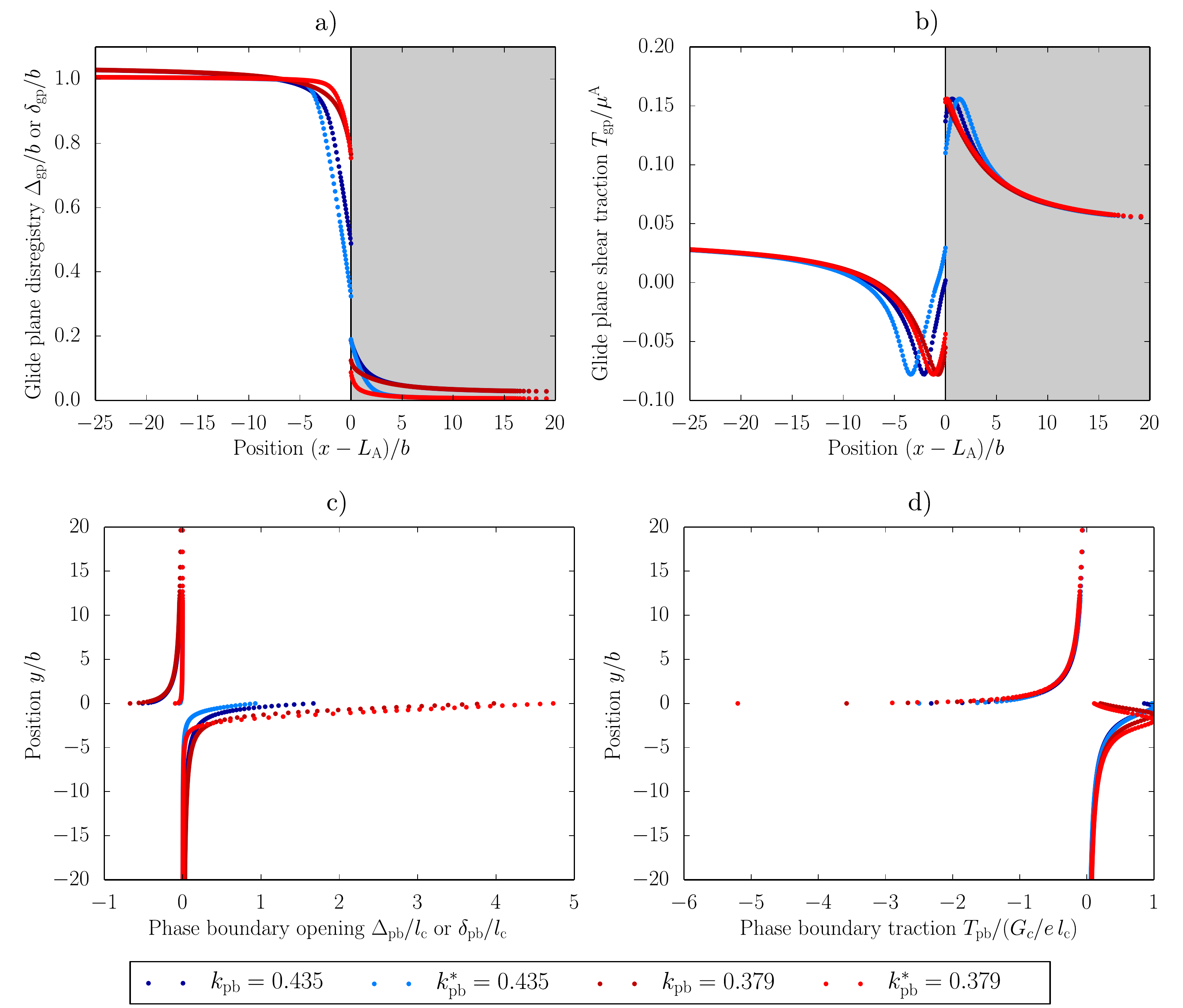}
	\caption{Model response for a single dislocation interacting with a decohering phase boundary with toughness factors $k_{\mathrm{pb}}=k_{\mathrm{pb}}^*\in\left\{0.379, 0.435\right\}$ at $\tau=0.04\mu^{\mathrm{A}}$: (a) disregistry profile $\Delta_{\mathrm{gp}}$ for the unreduced ($\psi_{\mathrm{gp}}$, $\psi_{\mathrm{pb}}$) and $\delta_{\mathrm{gp}}$ for the reduced potentials ($\psi_{\mathrm{gp}}^*$, $\psi_{\mathrm{pb}}^*$) and (b) glide plane traction $T_{\mathrm{gp}}=T_{\mathrm{gp}}^*$. (c) Opening profile $\Delta_{\mathrm{pb}}$ for the unreduced ($\psi_{\mathrm{gp}}$, $\psi_{\mathrm{pb}}$) and $\delta_{\mathrm{pb}}$ for the reduced potentials ($\psi_{\mathrm{gp}}^*$, $\psi_{\mathrm{pb}}^*$) and (d) phase boundary traction $T_{\mathrm{pb}}=T_{\mathrm{pb}}^*$.}
	\label{fig:gp-pb-Rice_comp-inc-6}
\end{figure}

Under an increased externally applied shear load, pushing the dislocation closer to the phase boundary, the impact of the potential reduction grows. This is illustrated for $\tau=0.04\,\mu^{\mathrm{A}}$ in Figure \ref{fig:gp-pb-Rice_comp-inc-6}a-b by the disregistry profiles and the glide plane tractions, and in Figure \ref{fig:gp-pb-Rice_comp-inc-6}c-d by the opening profiles and the phase boundary tractions. Here, two dominant influences of the reduced  potential are visible, as follows. For the tougher interface, $k_{\mathrm{pb}} = k_{\mathrm{pb}}^*=0.435$, the potential reduction leads to a lower phase boundary opening of $\delta_{\mathrm{pb}}(0^-)=0.93\,l_{\mathrm{c}}$ (versus $\Delta_{\mathrm{pb}}(0^-)=1.67\,l_{\mathrm{c}}$). For the weaker interface, $k_{\mathrm{pb}} = k_{\mathrm{pb}}^*=0.379$, it has the opposite effect and enhances the opening to $\delta_{\mathrm{pb}}(0^-)=4.73\,l_{\mathrm{c}}$ ($\Delta_{\mathrm{pb}}(0^-)=3.97\,l_{\mathrm{c}}$). These different behaviours between $k_{\mathrm{pb}}$ and $k_{\mathrm{pb}}^*$ stem from the highly non-linear interaction between the phase boundary opening and the bulk relaxation. Influential contributions are the initial phase boundary compliance, the softening behaviour of the phase boundary, as well as the difference in the glide plane potential (cf. Figures \ref{fig:Traction-profiles}, \ref{fig:normal-stress-tied-vs-pos_disreg}). In terms of dislocation transmission, the reduced potentials entail only a minor decrease of the external transmission stress for $k_{\mathrm{pb}}=k_{\mathrm{pb}}^*=0.435$ with $\tau_{\mathrm{trans}}^*= 1.07\,\tau_{\mathrm{trans}}^{\infty}$ ($\tau_{\mathrm{trans}} = 1.16\,\tau_{\mathrm{trans}}^{\infty}$). As observed earlier for $k_{\mathrm{pb}}=0.379$ (see Section \ref{sect:disl-vs-pb-wo_Rice}), no dislocation transmission is triggered for $k_{\mathrm{pb}}^*=0.379$ below an externally applied shear load of $\tau=\bar{\tau}$.\par
Note that the difference in model response strongly depends on the phase contrast $k_{\mathrm{m}}$ and the toughness factor $k_{\mathrm{pb}}$. While a higher value of $k_{\mathrm{m}}$ leads to enhanced dislocation obstruction \cite{Bormann2018}, $k_{\mathrm{pb}}$ sets the compliance of the phase boundary. Thus, with increasing $k_{\mathrm{pb}}$ (lower compliance) the influence of the potential reduction diminishes.
%
%
\subsubsection{Dislocation pile-up}
\begin{figure}[htbp]
	\centering
	\includegraphics[width=1.0\linewidth]{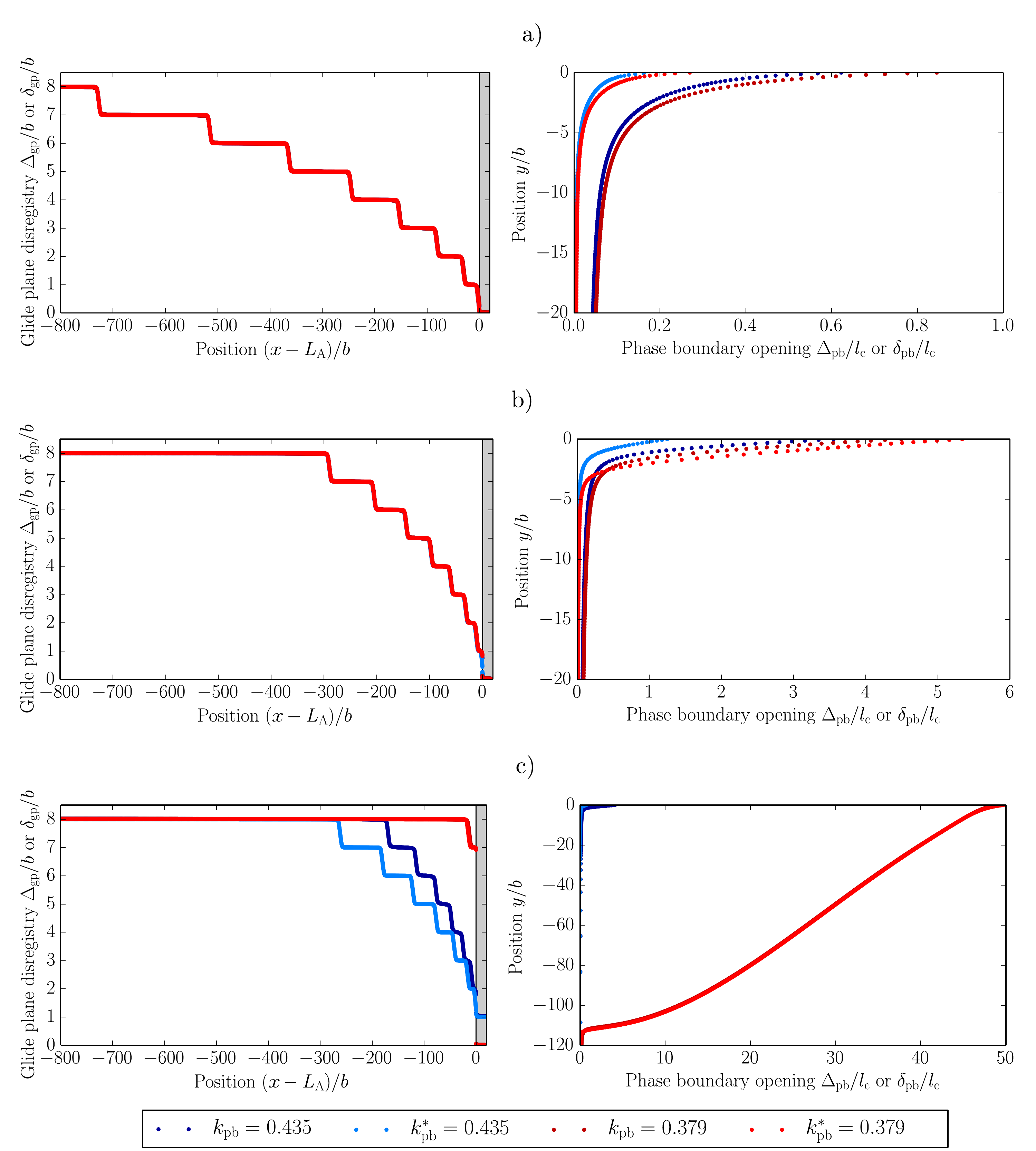}
	\caption{Disregistry profiles $\Delta_{\mathrm{gp}}$, $\delta_{\mathrm{gp}}$ and opening profiles $\Delta_{\mathrm{pb}}$, $\delta_{\mathrm{pb}}$ for an 8 dislocation pile-up system and a decohering phase boundary with the phase contrast $k_{\mathrm{m}}=2$ and toughness factors $k_{\mathrm{pb}}=k_{\mathrm{pb}}^*\in\left\{0.379,0.435\right\}$ under different externally applied shear loads $\tau$: (a) at $\tau=0.0054\,\mu^{\mathrm{A}}$, (b) at $\tau=0.0118\,\mu^{\mathrm{A}}$ and (c) after crack nucleation at $\tau=0.0182\mu^{\mathrm{A}}$ and $\tau^*=0.0174\mu^{\mathrm{A}}$, and after transmission at $\tau=0.0181\mu^{\mathrm{A}}$ and $\tau^*=0.0124\mu^{\mathrm{A}}$. Note the different scale for the opening.}
	\label{fig:Disregistry-8pu-Rice_comp}
\end{figure}

To demonstrate the influence of the potential reduction under the presence of multiple dislocations, an 8-dislocation pile-up system is considered. Results show only a negligible influence on the dislocation position before transmission or decohesion is triggered. This is illustrated in Figure \ref{fig:Disregistry-8pu-Rice_comp} in terms of the disregistry and opening profiles at different externally applied shear loads. Similar to the single dislocation case, $k_{\mathrm{pb}}^*=0.379$ evokes the largest and $k_{\mathrm{pb}}^*=0.435$ the smallest opening for the leading dislocation situated at the phase boundary. \par
With increasing externally applied shear load, dislocation transmission or phase boundary decohesion occurs. For $k_{\mathrm{pb}}^* = 0.379$ the reduction of the potential causes only a minor decrease of the external shear load causing decohesion at $\tau_{\mathrm{dec}}^*\approx 1.45\,\tau_{\mathrm{trans}}^{\infty}$ ($\tau_{\mathrm{dec}} = 1.52\,\tau_{\mathrm{trans}}^{\infty}$). No significant difference in phase boundary opening behaviour is observed. For $k_{\mathrm{pb}}^* = 0.435$, the external transmission stress is strongly affected and decreases to $\tau_{\mathrm{trans}}^*\approx 1.03\,\tau_{\mathrm{trans}}^{\infty}$ ($\tau_{\mathrm{trans}} = 1.51\,\tau_{\mathrm{trans}}^{\infty}$). For small and large toughness factors $k_{\mathrm{pb}}$, the influence of the potential reduction on transmission or decohesion is expected to diminish. \par
For the selected toughness factors, the reduced  potentials do not yield a change of mechanism (transmission or decohesion), nor a significantly different phase boundary opening in case of decohesion. However, there might be configurations where $k_{\mathrm{pb}}^*$ and $k_{\mathrm{pb}}$ do not only show quantitative but also qualitative differences, i.e. a damage of mechanism.
%
%
%
%
\section{Dislocation--phase boundary interaction with reduced potentials}
\label{sect:trans_vs_dec}
In this section, the interplay of dislocations with a decohering phase boundaries is studied in detail. Goal of this study is the assessment of the specific influence of the phase contrast $k_{\mathrm{m}}$ and the phase boundary toughness factor $k_{\mathrm{pb}}^*$, and hence the phase boundary toughness, on the competition between dislocation transmission and crack nucleation, and on the resulting crack length. For this purpose an 8-dislocation pile-up system is considered with model and material settings as specified in Section \ref{sect:Parameters_set}. First the general model evolution is explained in detail for transmission and crack nucleation. Subsequently, a parameter study is performed to assess in detail the influence of $k_{\mathrm{m}}$ and $k_{\mathrm{pb}}^*$ on the triggered mechanism (transmission or crack nucleation) and the respective evolution process. Finally, the influence of the chosen parameters on the resulting crack length is presented.
%
\subsection{General model evolution for transmission and crack nucleation}
Consider first the earlier discussed cases with phase contrast $k_{\mathrm{m}}=2$ and toughness factors $k_{\mathrm{pb}}^*\in\left\{0.379,0.435\right\}$ (cf. Figure \ref{fig:Disregistry-8pu-Rice_comp}). During the course of transmission (for $k_{\mathrm{pb}}^*=0.435$) the disregistry at the phase boundary, $x=L_{\mathrm{A}}$, evolves from initially $\delta_{\mathrm{gp}}=0$ (defect and stress free) to $\delta_{\mathrm{gp}}>b$ (transmitted dislocation). Temporarily, the phase boundary opens up, leading to a disregistry jump across $\Gamma_{\mathrm{pb}}$ with $\delta_{\mathrm{gp}}^{\mathrm{A}}>\delta_{\mathrm{gp}}^{\mathrm{B}}$, where $\delta_{\mathrm{gp}}^{\mathrm{A}}$ and $\delta_{\mathrm{gp}}^{\mathrm{B}}$ denote the disregistries at $x^{\mathrm{A}}\in\left\{x\in\Gamma_{\mathrm{gp}}^{\mathrm{A}}|x=L_{\mathrm{A}}\right\}$ and $x^{\mathrm{B}}\in\left\{x\in\Gamma_{\mathrm{gp}}^{\mathrm{B}}|x=L_{\mathrm{A}}\right\}$, respectively. 
In the case of crack nucleation (for $k_{\mathrm{pb}}^*=0.379$), the pile-up configuration evolves initially in a similar manner. The phase boundary opening, however, is somewhat more pronounced, leading to the absorption of the first dislocation into the phase boundary before, ultimately, crack nucleation is triggered. \par
\begin{figure}[htbp]
	\centering
	\includegraphics[width=1.0\linewidth]{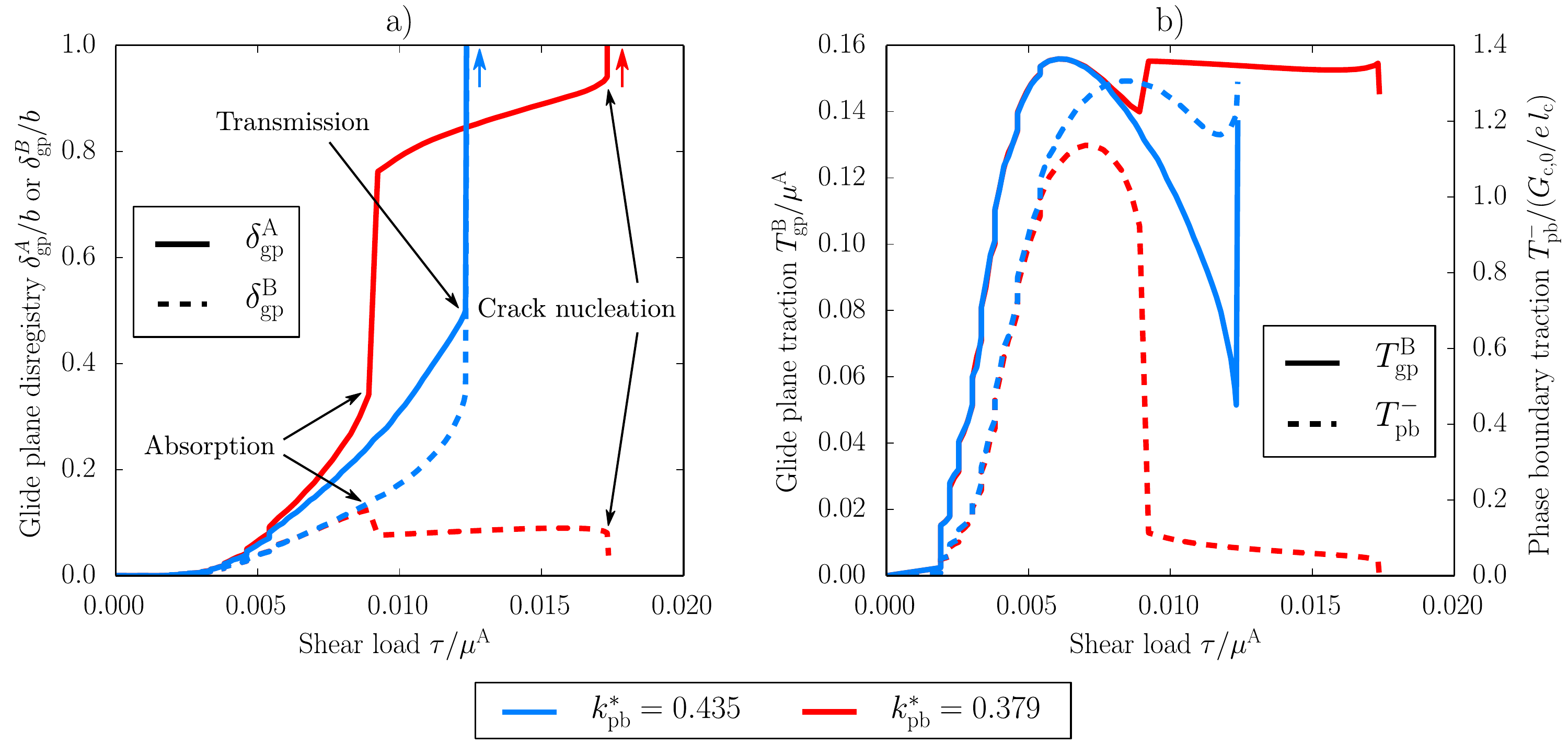}
	\caption{Model evolution for an 8-dislocation pile-up system up to crack nucleation or transmission with the phase contrast $k_{\mathrm{m}}=2$ and toughness factors $k_{\mathrm{pb}}^*\in\left\{0.379, 0.435\right\}$ as a function of the externally applied shear load $\tau$: (a) disregistries $\delta_{\mathrm{gp}}^{\mathrm{A}}$ at $\left\{x\in\Gamma_{\mathrm{gp}}^A|x=L_{\mathrm{A}}\right\}$, and $\delta_{\mathrm{gp}}^{\mathrm{B}}$ at $\left\{x\in\Gamma_{\mathrm{gp}}^B|x=L_{\mathrm{A}}\right\}$ and (b) glide plane traction $T_{\mathrm{gp}}^{\mathrm{B}}$ at $\left\{x\in\Gamma_{\mathrm{gp}}^B|x=L_{\mathrm{A}}\right\}$ and phase boundary traction $T_{\mathrm{pb}}^-$ at $y=0^-$.}
	\label{fig:mech_study}
\end{figure}

The corresponding evolutions of the disregistries $\delta_{\mathrm{gp}}^{\mathrm{A}}$ and $\delta_{\mathrm{gp}}^{\mathrm{B}}$ for $k_{\mathrm{pb}}^*\in\left\{0.379,0.435\right\}$ are illustrated in Figure \ref{fig:mech_study}a as a function of the externally applied shear load $\tau$. Due to the negligible compression above the glide plane, the disregistry jump practically equals the phase boundary opening: $\delta_{\mathrm{pb}}^-=\delta_{\mathrm{pb}}(y=0_-)\approx \delta_{\mathrm{gp}}^{\mathrm{A}} - \delta_{\mathrm{gp}}^{\mathrm{B}}$. During the evolution of the system with $k_{\mathrm{pb}}^*=0.379$, two jumps in the disregistry are apparent. These are characteristics for absorption of the leading dislocation (first jump, at $\tau/\mu^{\mathrm{A}}\approx 0.009$) and crack nucleation (second jump, at $\tau/\mu^{\mathrm{A}}\approx 0.017$). For $k_{\mathrm{pb}}^*=0.435$ the strong increase in disregistry beyond $\delta_{\mathrm{gp}} = b$ indicates the point of dislocation transmission and the migration of the next dislocation in the pile-up to the boundary. \par
The different model responses suggest that the first dislocation's absorption leads to a bifurcation, where the model either progresses further towards dislocation transmission or diverts towards crack nucleation. As the dislocation is being absorbed, the surrounding bulk relaxes, increasing the barrier against dislocation transmission (cf. Section \ref{sect:Unreduced_Results}). After the leading dislocation is being absorbed, the externally applied shear load needs to be increased further to nucleate a crack.\par
To obtain a better insight into the underlying mechanics of the system, Figure \ref{fig:mech_study}b plots the evolution of the glide plane traction of Phase~$\mathrm{B}$ $T_{\mathrm{gp}}^B=T_{\mathrm{gp}}(x=L_{\mathrm{A}})$ and of the phase boundary traction $T_{\mathrm{pb}}^-=T_{\mathrm{pb}}(y=0^-)$. These evolution profiles reflect the influence of the successive nucleation of dislocations (traction jumps for $\tau<0.006\,\mu^{\mathrm{A}}$), dislocation transmission (last traction jump for $k_{\mathrm{pb}}^*=0.435$), as well as dislocation absorption and crack nucleation (last two traction jumps for $k_{\mathrm{pb}}^*=0.379$). The initially similar model response for both toughness factors corresponds to a comparable phase boundary behaviour in the early stages of the model evolution. With increasing $\tau$, the tractions begin to diverge, highlighting the strong influence of the phase boundary toughness. \par
For $k_{\mathrm{pb}}^*=0.435$, where transmission is triggered, $T_{\mathrm{gp}}^{\mathrm{B}}$ decreases after reaching the traction amplitude $\max\left\{T_{\mathrm{gp}}\right\}$ of Phase~$\mathrm{B}$, corresponding to the increase of $\delta_{\mathrm{gp}}^{\mathrm{B}}$. Simultaneously, the phase boundary opens up beyond the peak traction. With the ongoing transmission process, the phase boundary softening is in a constant stable equilibrium with the related bulk relaxation. Ultimately, the transmission process is advanced to such an extent, that the dislocation induced traction, exerted on the phase boundary, begins to decrease and the phase boundary opening process reverses -- the dislocation is being transmitted. \par
For the weaker interface, on the contrary, the peak traction is reached at an earlier stage, since a lower dislocation induced traction and hence less pile-up compression is needed. With the continuation of the evolution, the leading dislocation is pushed further towards the phase boundary, leading to an increase in phase boundary opening. Eventually, a critical point is reached where the phase boundary softening is not in stable equilibrium anymore with the related bulk relaxation. This results in the leading dislocation being absorbed instantly into the phase boundary.\par
It thus can be anticipated that there exists a toughness factor $k_{\mathrm{pb,s}}^*$ at which the mechanism changes from crack nucleation to transmission.\par
%
%
\subsection{Parameter study on dislocation transmission vs. crack nucleation}
For a detailed study of the competition between dislocation transmission and crack nucleation we continue to consider the 8-dislocation pile-up system, but vary the phase contrast $k_{\mathrm{m}}$ and toughness factor $k_{\mathrm{pb}}^*$. An equivalent study for a 4-dislocation pile-up system showed a similar qualitative behaviour and is therefore not included.\par
\begin{figure}[htbp]
	\centering
	\includegraphics[width=1.0\linewidth]{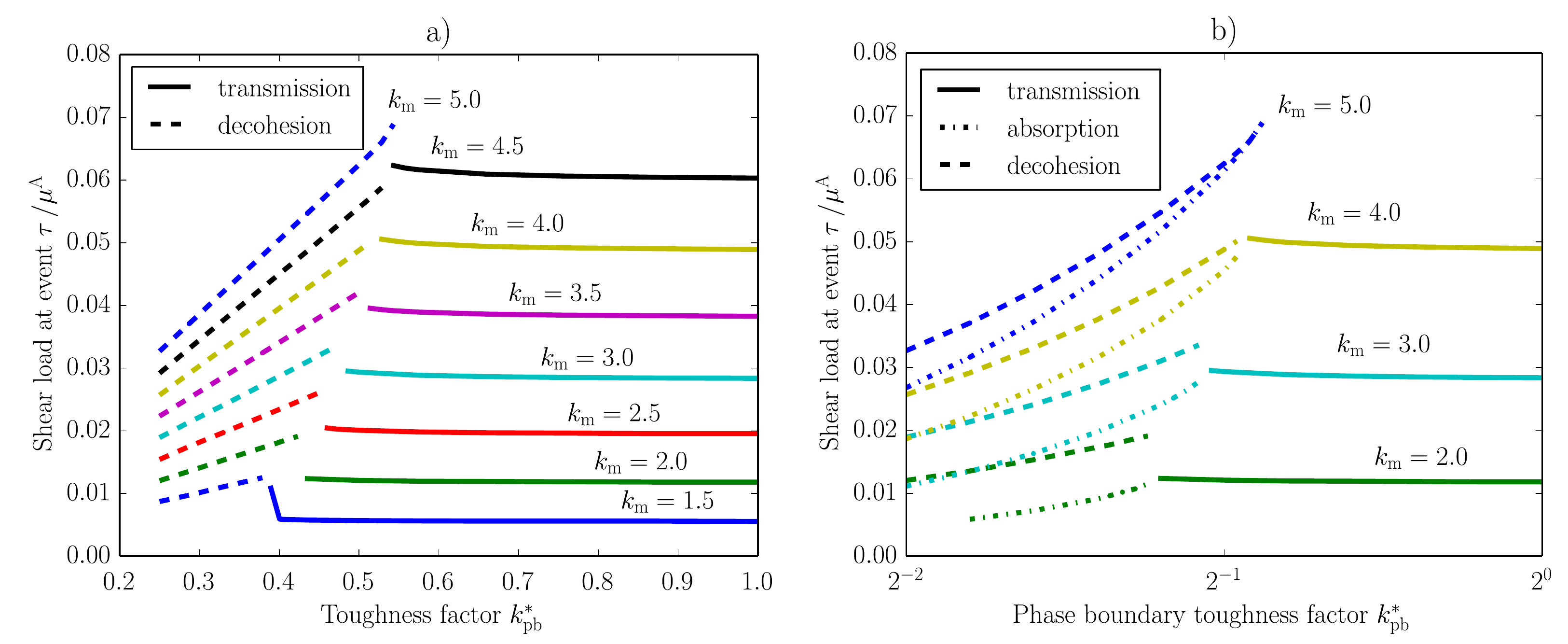}
	\caption{Model response for an 8-dislocation pile-up system as a function of the toughness factor $k_{\mathrm{pb}}^*$ for various phase contrasts $k_{\mathrm{m}}$: (a) externally applied shear load $\tau$ at crack nucleation or transmission and (b) externally applied shear load $\tau$ at dislocation absorption, crack nucleation or transmission for selected $k_{\mathrm{m}}$.}
	\label{fig:event-N=8}
\end{figure}
\begin{figure}[htbp]
	\centering
	\includegraphics[width=0.5\linewidth]{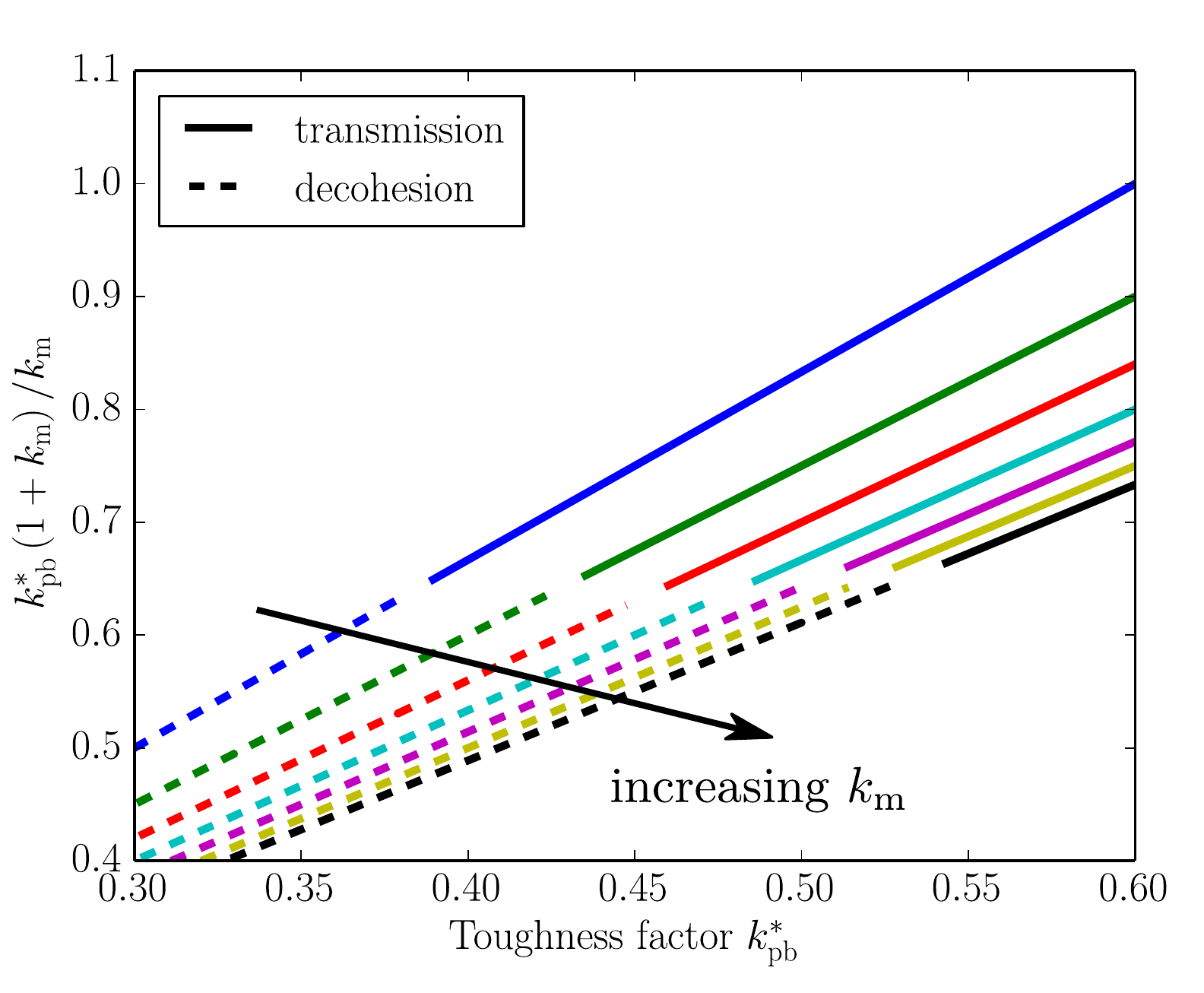}
	\caption{Ratio $k_{\mathrm{pb}}^*\left(1+k_{\mathrm{m}}\right)/k_{\mathrm{m}}$ for the illustration of the transition of the model from crack nucleation to transmission, representative for $G_{\mathrm{c}}/\psi_{\mathrm{gp}}^{\mathrm{B}*}(\delta_{\mathrm{gp}}=b/2)$ and $\max\left\{T_{\mathrm{pb}}\right\}/\max\left\{T_{\mathrm{gp}}^{\mathrm{B}}\right\}$.}
	\label{fig:event-ratio}
\end{figure}

The influence of the phase contrast $k_{\mathrm{m}}$ and the toughness factor $k_{\mathrm{pb}}^*$ on the model response is presented in Figure \ref{fig:event-N=8}. Plotted is the externally applied shear load $\tau$ at transmission (solid line), at dislocation absorption (dash-dotted line) or at crack nucleation (dashed line). Under the maximum applied shear load of $\tau/\mu^{\mathrm{A}}=0.07$ no event is triggered for $k_{\mathrm{m}}=5 $ and $k_{\mathrm{pb}}^*$ greater than approximately $0.54$. This comparison demonstrates the complex interplay between absorption, crack nucleation and transmission during the approach of dislocations towards phase boundaries (see Figure \ref{fig:event-N=8}b). Three changes of mechanism are noticeable. First, for $k_{\mathrm{m}}=5$ and toughness factors
$0.529\le k_{\mathrm{pb}}^*\le 0.514$ dislocation absorption invokes immediately nucleation of a crack and does not require an increase in shear load $\tau$. Second, for $k_{\mathrm{m}}=1.5$ and $k_{\mathrm{pb}}^*=0.390$, although the dislocation induced tractions lead to an opening which triggers the absorption of the leading dislocation, the tractions of the remaining pile-up do not suffice to trigger the nucleation of a crack. Eventually, the leading, absorbed, dislocation is being transmitted instead. Third, a change of mechanism from crack nucleation to transmission is observed at toughness factors around $k_{\mathrm{pb}}^*\approx 0.64\,k_{\mathrm{m}}/\left(1+k_{\mathrm{m}}\right)$, as illustrated in Figure \ref{fig:event-ratio}. This value is representative for the ratios $G_{\mathrm{c}}/\psi_{\mathrm{gp}}^{\mathrm{B}*}(\delta_{\mathrm{gp}}=b/2) \approx 2.23$ and $\max\left\{T_{\mathrm{pb}}\right\}/\max\left\{T_{\mathrm{gp}}^{\mathrm{B}}\right\}\approx 1.58$. This constant ratio shows that the relative height of the energy barriers associated with decohesion ($G_{\mathrm{c}}$) and transmission ($\psi_{\mathrm{gp}}^{\mathrm{B}*}$) is the decisive factor in the outcome of the competition between transmission and crack nucleation. \par
Below, the specific influences of $k_{\mathrm{m}}$ and $k_{\mathrm{pb}}^*$ on the evolution process will be discussed in detail to elaborate on the specific trends observed in Figure \ref{fig:event-N=8}.\par
%
%
\subsubsection{Influence of the toughness factor $k_{\mathrm{pb}}^*$}
\begin{figure}[htbp]
	\centering
	\includegraphics[width=1.0\linewidth]{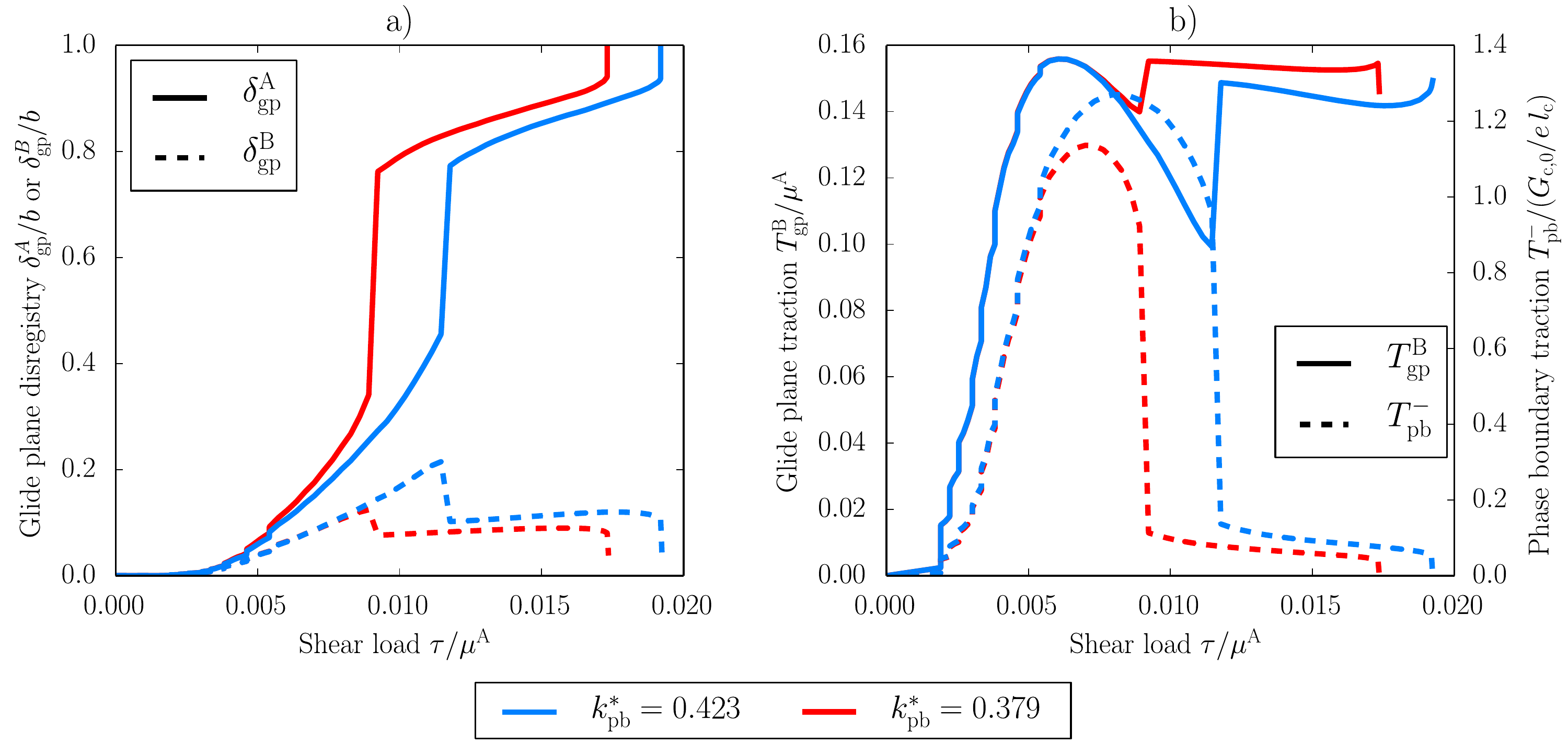}
	\caption{Model evolution for an 8-dislocation pile-up system up to crack nucleation with the phase contrast $k_{\mathrm{m}}=2$ and toughness factors $k_{\mathrm{pb}}^*\in\left\{0.379, 0.423\right\}$ as a function of the externally applied shear load $\tau$: (a) disregistries $\delta_{\mathrm{gp}}^{\mathrm{A}}$ at $\left\{x\in\Gamma_{\mathrm{gp}}^A|x=L_{\mathrm{A}}\right\}$, and $\delta_{\mathrm{gp}}^{\mathrm{B}}$ at $\left\{x\in\Gamma_{\mathrm{gp}}^B|x=L_{\mathrm{A}}\right\}$ and (b) glide plane traction $T_{\mathrm{gp}}^{\mathrm{B}}$ at $\left\{x\in\Gamma_{\mathrm{gp}}^B|x=L_{\mathrm{A}}\right\}$ and phase boundary traction $T_{\mathrm{pb}}^-$ at $y=0^-$.}
	\label{fig:k_pb_influence_dec}
\end{figure}

To assess the influence of the phase boundary toughness (and strength) only on crack nucleation and dislocation transmission, consider the constant phase contrast $k_{\mathrm{m}}=2$ under varying toughness factor $k_{\mathrm{pb}}^*$. Results show that an increasing $k_{\mathrm{pb}}^*$ shifts the points of absorption ($\tau_{\mathrm{abs}}$) and crack nucleation ($\tau_{\mathrm{dec}}$) towards higher externally applied shear loads, as illustrated by the disregistry evolution of $\delta_{\mathrm{gp}}^{\mathrm{A}}$ and $\delta_{\mathrm{gp}}^{\mathrm{B}}$ in Figure \ref{fig:k_pb_influence_dec}a for $k_{\mathrm{pb}}^*\in\left\{0.379, 0.423\right\}$. The evolutions of the corresponding tractions $T_{\mathrm{gp}}^{\mathrm{B}}$ and $T_{\mathrm{pb}}^-$ illustrate the origin of the delayed absorption for stronger interfaces. As $k_{\mathrm{pb}}^*$ is increased, the toughness and strength of the phase boundary increase likewise. Thus, higher dislocation induced stresses, and hence a larger pile-up compression is required for dislocation absorption. Equally, to nucleate a crack in a stronger phase boundary (after absorption), the pile-up needs to be compressed more. In this context, a stronger increase of $\tau_{\mathrm{abs}}$ than of $\tau_{\mathrm{dec}}$ is observed. Eventually, under sufficiently large $k_{\mathrm{pb}}^*$ ($\ge k_{\mathrm{pb,s}}^*$), the dislocation induced tractions no longer suffice to trigger dislocation absorption. The leading dislocation is being transmitted instead.\par
\begin{figure}[htbp]
	\centering
	\includegraphics[width=1.0\linewidth]{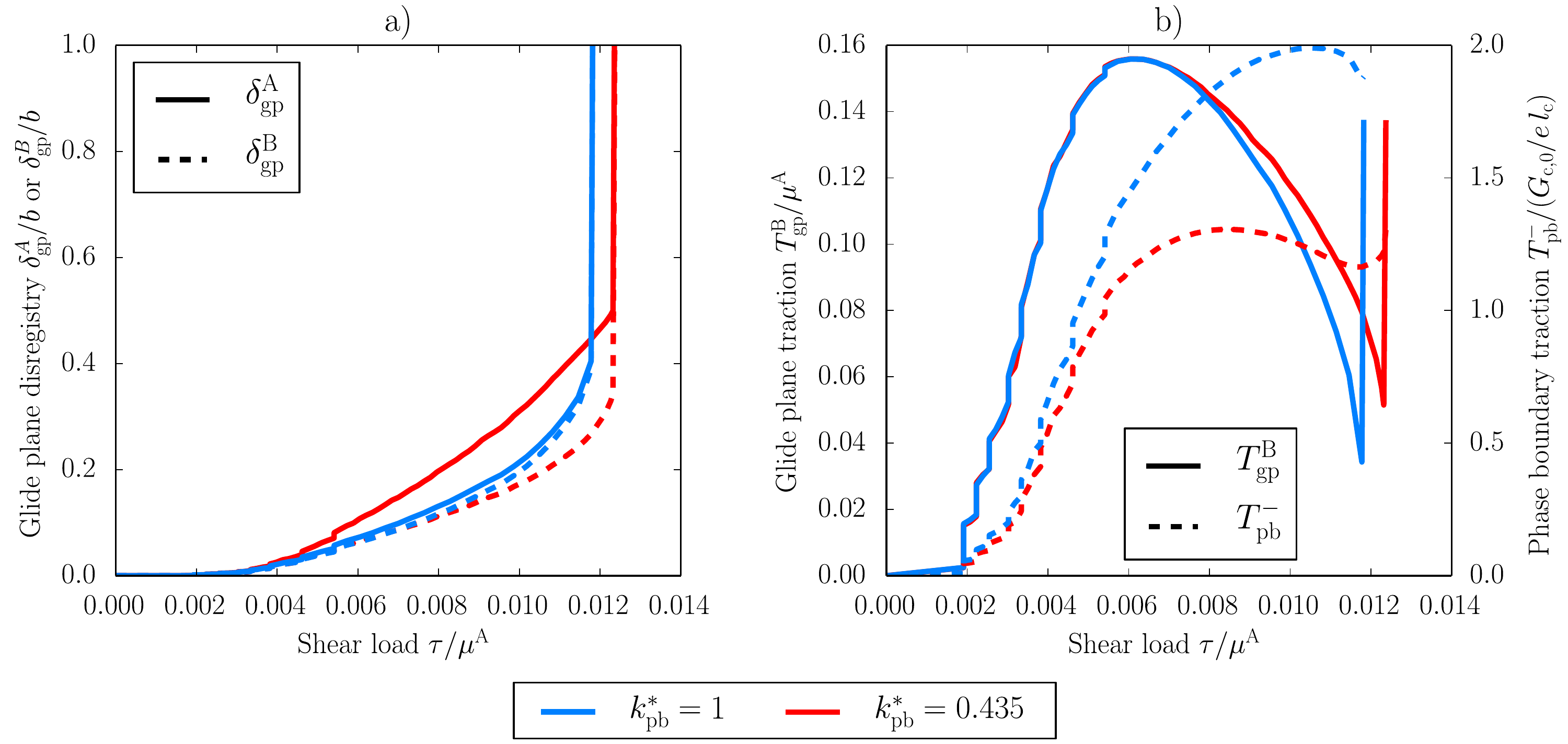}
	\caption{Model evolution for an 8-dislocation pile-up system up to transmission with the phase contrast $k_{\mathrm{m}}=2$ and toughness factors $k_{\mathrm{pb}}^*\in\left\{0.435, 1\right\}$ as a function of the externally applied shear load $\tau$: (a) disregistries $\delta_{\mathrm{gp}}^{\mathrm{A}}$ at $\left\{x\in\Gamma_{\mathrm{gp}}^A|x=L_{\mathrm{A}}\right\}$, and $\delta_{\mathrm{gp}}^{\mathrm{B}}$ at $\left\{x\in\Gamma_{\mathrm{gp}}^B|x=L_{\mathrm{A}}\right\}$ and (b) glide plane traction $T_{\mathrm{gp}}^{\mathrm{B}}$ at $\left\{x\in\Gamma_{\mathrm{gp}}^B|x=L_{\mathrm{A}}\right\}$ and phase boundary traction $T_{\mathrm{pb}}^-$ at $y=0^-$.}
	\label{fig:k_pb_influence_trans}
\end{figure}

Although a further increase in $k_{\mathrm{pb}}^*$ reduces the phase boundary opening, and in turn the bulk relaxation, for $k_{\mathrm{m}}=2$ it has only a marginal influence on the transmission behaviour and the external transmission stress $\tau_{\mathrm{trans}}$, as observed by the plateau in Figure \ref{fig:event-N=8}. The corresponding disregistry evolutions $\delta_{\mathrm{gp}}^{\mathrm{B}}$ and $\delta_{\mathrm{gp}}^{\mathrm{A}}$ are illustrated in Figure \ref{fig:k_pb_influence_trans} for $k_{\mathrm{pb}}^*\in\left\{0.435, 1\right\}$ together with the evolution of the tractions $T_{\mathrm{gp}}^{\mathrm{B}}$ and $T_{\mathrm{pb}}^-$. This demonstrates that, despite a relatively large difference in the opening behaviour, no significant difference in the transmission process is present.
%
%
\subsubsection{Influence of the phase contrast $k_{\mathrm{m}}$}
Consider next the impact of $k_{\mathrm{m}}$ on the externally applied shear load to trigger absorption, crack nucleation or transmission. Within the present PN-CZ model, the phase contrast $k_{\mathrm{m}}$ has a three-fold influence on the model response. First, it evokes a repulsive image stress on dislocations and is thus a strong source of dislocation obstruction. Second, it affects via the toughness $G_{\mathrm{c}}\propto k_{\mathrm{pb}}^*(1+k_{\mathrm{m}})$ the phase boundary opening behaviour. Third, it defines the maximum dislocation induced traction on the phase boundary. By reducing the influence of the phase boundary opening, with $k_{\mathrm{pb}}^*=1$, the impact on the repulsive image stresses is determined. In that context, results reveal a non-linear correlation between $k_{\mathrm{m}}$ and the repulsive image stresses, reflected by the non-linear increase of the external transmission stress $\tau_{\mathrm{trans}}$ (cf. Figure \ref{fig:event-N=8}). \par
\begin{figure}[htbp]
	\centering
	\includegraphics[width=1.0\linewidth]{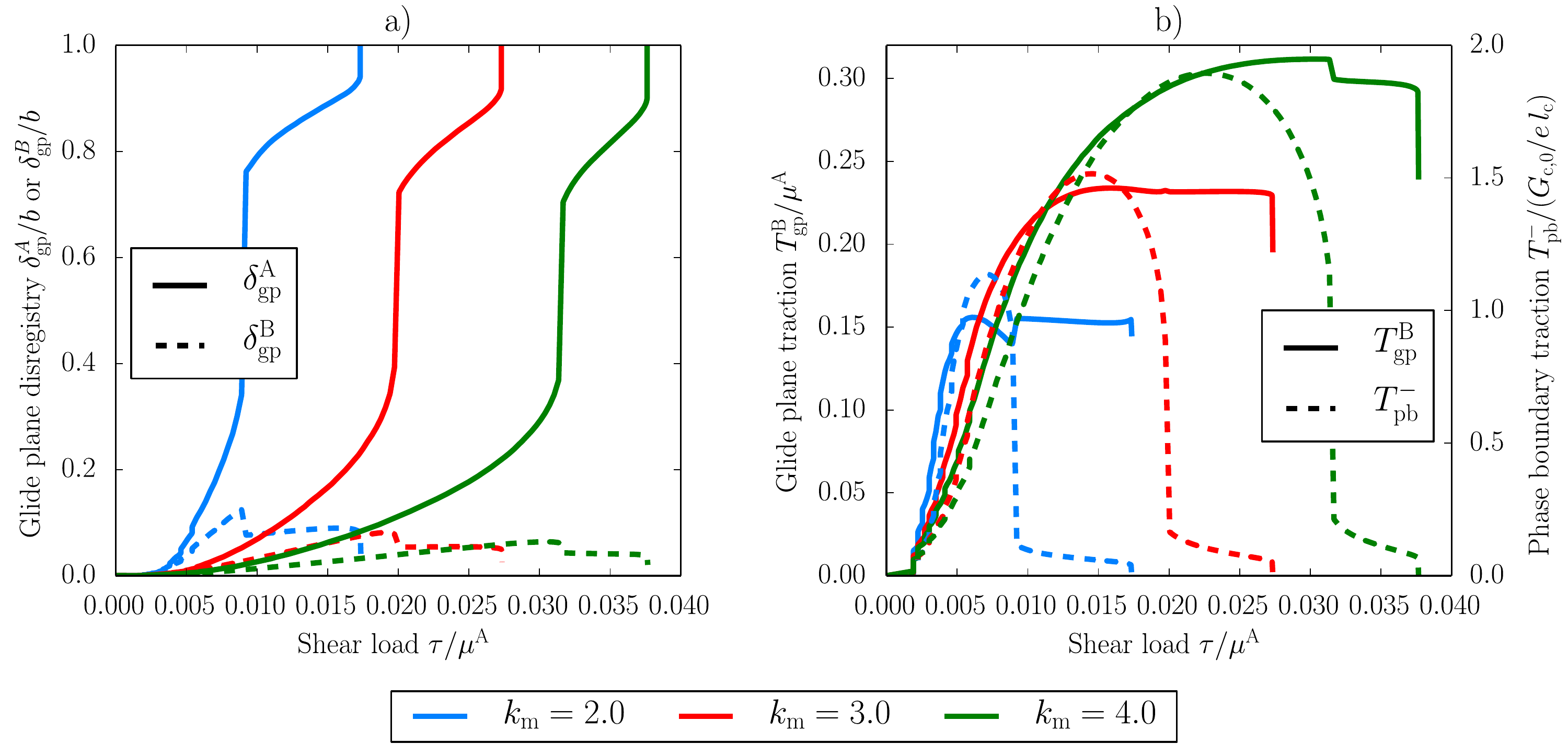}
	\caption{Model evolution for an 8-dislocation pile-up system up to decohesion with the phase contrasts $k_{\mathrm{m}}\in\left\{2,3,4\right\}$ and a constant toughness factor of $k_{\mathrm{pb}}^*=0.379$ as a function of the externally applied shear load $\tau$: (a) disregistries $\delta_{\mathrm{gp}}^{\mathrm{A}}$ at $\left\{x\in\Gamma_{\mathrm{gp}}^A|x=L_{\mathrm{A}}\right\}$, and $\delta_{\mathrm{gp}}^{\mathrm{B}}$ at $\left\{x\in\Gamma_{\mathrm{gp}}^B|x=L_{\mathrm{A}}\right\}$ and (b) glide plane traction $T_{\mathrm{gp}}^{\mathrm{B}}$ at $\left\{x\in\Gamma_{\mathrm{gp}}^B|x=L_{\mathrm{A}}\right\}$ and phase boundary traction $T_{\mathrm{pb}}^-$ at $y=0^-$.}
	\label{fig:const_k_pb-dec}
\end{figure}

To assess the influence of $k_{\mathrm{m}}$ on crack nucleation, $k_{\mathrm{pb}}^*$ is kept constant under varying $k_{\mathrm{m}}$. Figure \ref{fig:const_k_pb-dec} illustrates the model responses for $k_{\mathrm{pb}}^*=0.379$ and $k_{\mathrm{m}}\in\left\{2,3,4\right\}$ in terms of the evolutions of the disregistries $\delta_{\mathrm{gp}}^{\mathrm{A}}$ and $\delta_{\mathrm{gp}}^{\mathrm{B}}$, and the tractions $T_{\mathrm{gp}}^{\mathrm{B}}$ and $T_{\mathrm{pb}}^-$. The results show that with increasing $k_{\mathrm{m}}$ dislocation absorption is being triggered at an earlier stage of the transmission process (lower $\delta_{\mathrm{gp}}^{\mathrm{B}}$). As a consequence, the toughness factor $k_{\mathrm{pb}}^*$ can be increased further before the mechanism changes from crack nucleation to transmission, which explains the shift of $k_{\mathrm{pb,s}}^*$ for larger $k_{\mathrm{m}}$, as observed in Figure \ref{fig:event-N=8}. Similarly, with larger $k_{\mathrm{m}}$ less increase in the pile-up compression is required to ultimately trigger crack nucleation. \par
\begin{figure}[htbp]
	\centering
	\includegraphics[width=1.0\linewidth]{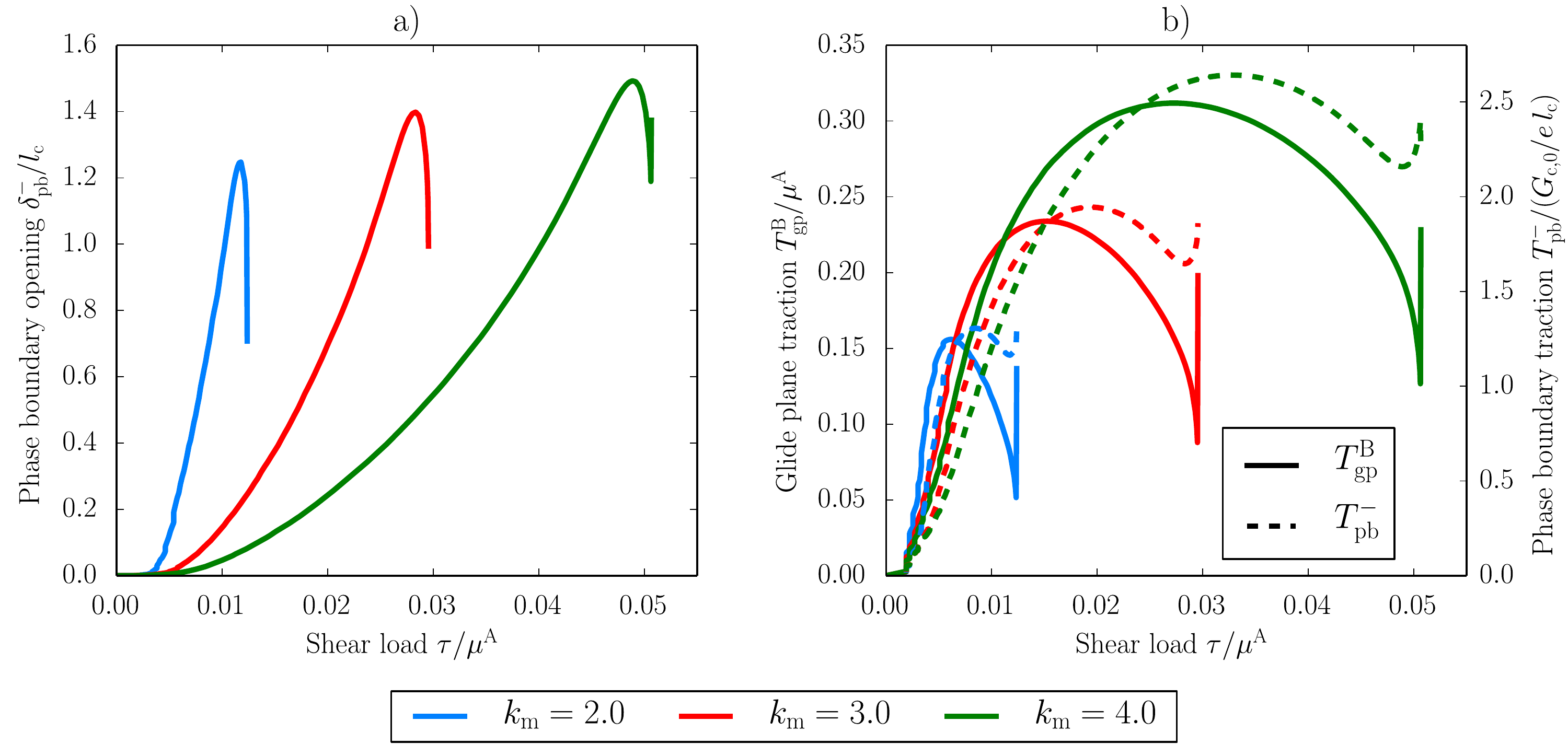}
	\caption{Model evolution for an 8-dislocation pile-up system up to transmission with the phase contrasts $k_{\mathrm{m}}=\{2,3,4\}$ and transmission evoking toughness factors $k_{\mathrm{pb}}^*=k_{\mathrm{pb,s}}^*$ as a function of the externally applied shear load $\tau$: (a) opening $\delta_{\mathrm{pb}}^-$ at $y=0^-$ and (b) glide plane traction $T_{\mathrm{gp}}^{\mathrm{B}}$ at $\left\{x\in\Gamma_{\mathrm{gp}}^B|x=L_{\mathrm{A}}\right\}$ and phase boundary traction $T_{\mathrm{pb}}^-$ at $y=0^-$.}
	\label{fig:k_pbs_trans}
\end{figure}

In terms of dislocation transmission, a growing impact of $k_{\mathrm{pb}}^*$ on $\tau_{\mathrm{trans}}$ is noticeable as $k_{\mathrm{m}}$ increases. This effect is related to larger phase boundary openings at $k_{\mathrm{pb,s}}^*$, as visualised in Figure \ref{fig:k_pbs_trans}a by the phase boundary openings $\delta_{\mathrm{pb}}^-$, which increase with increasing $k_{\mathrm{m}}$. The corresponding tractions $T_{\mathrm{gp}}^{\mathrm{B}}$ and $T_{\mathrm{pb}}^-$ are displayed for completeness in Figure \ref{fig:k_pbs_trans}b. In relation with the pronounced opening behaviour for larger $k_{\mathrm{m}}$, a higher bulk relaxation applies, which requires an increasingly larger external transmission stress $\tau_{\mathrm{trans}}$ than for the damage free boundary $k_{\mathrm{pb}}^*=\infty$. 
\subsection{Crack response}
For all cases where a crack is nucleated, an immediate crack growth is observed, with the absorption of, in addition to the leading dislocation, 6 dislocations into the phase boundary (cf. Figure \ref{fig:Disregistry-8pu-Rice_comp}c for $k_{\mathrm{m}}=2.0$ and $k_{\mathrm{pb}}^*=0.379$). Note that this observation is limited to the present 8-dislocation pile-up system. In a similar simulation of a 23-dislocation pile-up system with $k_{\mathrm{m}}=2$ and $k_{\mathrm{pb}}^*=0.379$, not shown here, an absorption of 18 dislocations was observed. \par
\begin{figure}[htbp]
	\centering
	\includegraphics[width=1.0\linewidth]{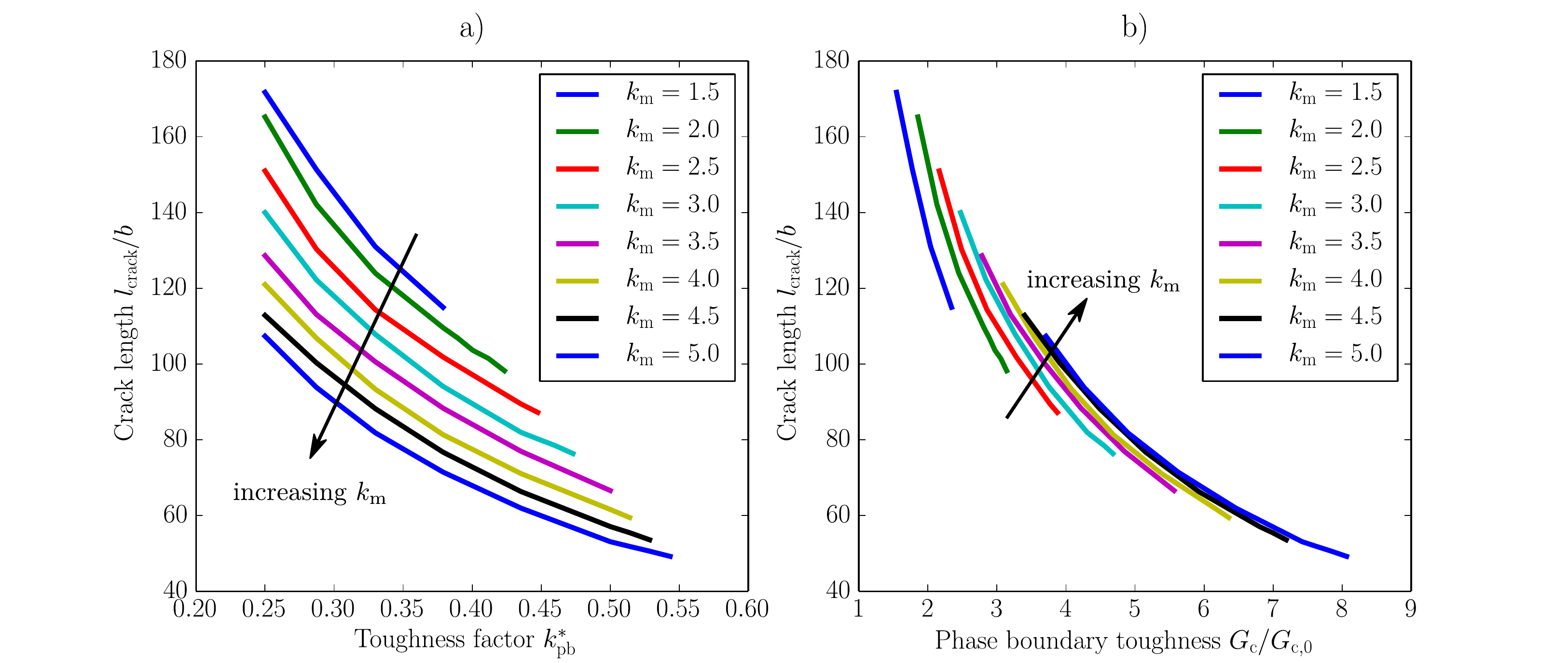}
	\caption{Crack length $l_{\mathrm{crack}}$ after decohesion for an 8-dislocation pile-up system under various phase contrasts $k_{\mathrm{m}}$: (a) as a function of the toughness factor $k_{\mathrm{pb}}^*$ and (b) as a function of the phase boundary toughness $G_{\mathrm{c}}$.}
	\label{fig:Crack_length}
\end{figure}

Although the 8-dislocation pile-up system in the case of crack nucleation always exhibits an equal number of 7 dislocations absorbed into the phase boundary, the specific model responses strongly differ in their crack opening behaviour. This is illustrated in Figure \ref{fig:Crack_length} in terms of crack length $l_{\mathrm{crack}}$ as a function of $k_{\mathrm{pb}}^*$ and as a function of the phase boundary toughness $G_{\mathrm{c}}$. Here, the crack length is defined as the distance between $y=0$ and the position of the crack tip, where $T_{\mathrm{pb}}<0.1\,\max\left\{T_{\mathrm{pb}}\right\}$ and $\delta_{\mathrm{pb}}>l_{\mathrm{c}}$. As expected, the crack length is directly dependent on the work of separation. Furthermore, a larger phase contrast $k_{\mathrm{m}}$ entails a closer dislocation position to the phase boundary as the dislocation induced tractions are larger, leading to a further increase in $l_{\mathrm{crack}}$ with $k_{\mathrm{m}}$. \par
%
%
%
%
\section{Conclusion}
\label{sect:discussion}
In this paper, a previously proposed Peierls--Nabarro finite element model \cite{Bormann2018} was complemented with a model that accounts for decohesion of a phase boundary, resulting in a Peierls--Nabarro cohesive zone (PN-CZ) model. Its total free energy is formulated on the basis of linear elastic strain energy density, a glide plane potential for dislocation behaviour and the cohesive phase boundary potential. It was shown that with the cohesive zone model along the phase boundary, a strong influence on the dislocation behaviour is introduced. Depending on the phase boundary toughness, either dislocation transmission or phase boundary decohesion may be triggered. However, the results demonstrated that atomistically calculated glide plane and phase boundary potentials may lead, when directly used in zero-thickness interfaces (as in the PN-CZ model), to a large quantitative deviation in the applied shear load, required for transmission. Accordingly, a linear elastic potential reduction was incorporated to restore physical consistency.\par 
With the reduced potentials, the interplay between dislocation transmission and phase boundary decohesion was studied. Subject of this study was the behaviour of an 8-dislocation (dipole) pile-up system for a varying phase contrast $k_{\mathrm{m}}$ (in elasticity and glide plane properties) and interface toughness $G_{\mathrm{c}}\propto k_{\mathrm{pb}}\left(1+k_{\mathrm{m}}\right)$. The toughness factor $k_{\mathrm{pb}}$ at which the mechanism changes from crack nucleation to transmission was identified as $k_{\mathrm{pb}}\approx 0.64\,k_{\mathrm{m}}/\left(1+k_{\mathrm{m}}\right)$. During the evolution of transmission and decohesion,
under an increasing externally applied shear load, there exists a bifurcation point where the model either progresses further towards dislocation transmission or towards phase boundary decohesion. This point is characterised by the absorption of the leading dislocation into the phase boundary as a results of the non-linear interaction between phase boundary opening and the bulk relaxation, and generally occurs well before the actual decohesion/transmission. For a fixed phase contrast and increasing interface toughness it was shown that the points of first dislocation absorption and crack nucleation shift unequally towards larger externally applied shear loads. For dislocation transmission, a minor decrease of the required external transmission stress was revealed for stronger interfaces due to the smaller phase boundary opening and bulk relaxation. Naturally, with larger phase contrast stronger repulsive image stresses are induced, leading to a larger barrier to dislocation transmission. Hence, to overcome the higher repulsive image stresses and to trigger dislocation transmission or crack nucleation, a greater pile-up compression under larger externally applied shear load is required. In this context, it was revealed that the toughness factor for which the mechanism changes from crack nucleation to transmission shifts to larger values for increasing $k_{\mathrm{m}}$. Again, the points of first dislocation absorption and crack nucleation shift unequally. This unequal shift leads to the convergence of both points, which tend to overlap for a high phase contrast where absorption of the leading dislocation leads to an immediate crack nucleation. As the dislocation induced normal stress increases with the phase contrast, the phase boundary needs to be increasingly stronger to trigger dislocation transmission instead of phase boundary decohesion. \par
In all cases of decohesion, an immediate crack propagation appears until all but one dislocation are absorbed. This phenomenon is limited to the studied 8-dislocation pile-up system, as for a similar 23-dislocation pile-up system 5 remaining dislocations after crack nucleation were observed. An analysis of the resulting crack length corresponding to the 8-dislocation pile-up showed a strong influence of the phase boundary toughness. Furthermore, as a result of the decreasing distance of the remaining dislocation to the phase boundary and the accordingly increasing dislocation induced stress, the crack length grows with the phase contrast.\par
The present study was performed for the case of an 8-dislocation pile-up system. Once the restriction on the number of dislocations is lifted, more stable dislocations may be generated. Thus, based on the increase of the external decohesion stress with larger phase contrasts, an increase in the number of nucleated dislocations before failure may be anticipated. \par
Here, the idealised case of a glide plane perpendicular to and continuous across a fully coherent phase boundary was considered. For more complex phase boundary structures however, different responses may be expected, including the toughness factor at which the mechanism changes. The presence of a phase boundary boundary structure gives rise to a local coherency stress field. Depending on its positioning with respect to the impinging glide plane, dislocation transmission may be either promoted or impeded. Hence, the interplay between dislocation transmission and crack nucleation may shift. Furthermore, if a crack nucleates in a region of low coherency, it propagation may be impeded in regions of high coherency, requiring an increase of the externally applied shear load for further crack propagation. All of these effects will be subject of future work.
%
%
%
%
\section*{Acknowledgements}
We would like to thank Franz Roters and Pratheek Shanthraj of the Max Planck Institute for Iron Research for useful discussions. This research is supported by Tata Steel Europe through the Materials innovation institute (M2i) and Netherlands Organisation for Scientific Research (NWO), under the grant number STW 13358 and M2i project number S22.2.1349a.
\section*{References}
%
%
%
%

\end{document}